\documentclass[12pt]{article}
\usepackage{amsmath}
\usepackage{amssymb}
\usepackage{txfonts} 

\newtheorem{deff}{Definition}

\newtheorem{proposition}{Proposition}
\newtheorem{example}{Example}
\newtheorem{theorem}{Theorem}
\newtheorem{lemma}{Lemma}

\newtheorem{cor}{Corollary}

\newcommand{\proof}{\vskip2mm \noindent {\bf Proof.}~}
\newcommand{\End}{\mathop{\mathrm{End}}}

\newcommand{\rt}{\rightarrow}

\newcommand{\mto}{\mapsto}
\newcommand{\na}{\nabla}

\newcommand\beqa {\begin{eqnarray}}
\newcommand\eeqa {\end{eqnarray}}
\newcommand\bqa {\begin{eqnarray}}
\newcommand\eqa {\end{eqnarray}}
\newcommand{\beq}{\begin{eqnarray}}
\newcommand{\beqn}{\begin{eqnarray}\nonumber}
\newcommand{\eeq}{\end{eqnarray}}
\newcommand{\be}{\begin{array}}
\newcommand{\ee}{\end{array}}
\newcommand{\s}{\sigma}

\newcommand\bea {\begin{eqnarray}}
\newcommand\eea {\end{eqnarray}}

 \newcommand{\pt}{\partial}

 \newcommand{\cP}{{\cal P}}
 
 \newcommand{\cA}{{\cal A}}
 \newcommand{\cD}{{\cal D}}
 \newcommand{\cO}{{\cal O}}

 \newcommand{\cL}{{\cal L}}

\newcommand{\R}{{\mathbb R}}

 \newcommand{\g}{{\mathfrak g}}

 \newcommand{\h}{{\mathfrak h}}

\newcommand{\md}{\mathrm{d}}

 \def\S{{\Sigma}}
\def\2{{\textstyle\frac{1}{2}}}
\def\ba{\begin{eqnarray}}
\def\ea{\end{eqnarray}}

  \def\CA{{\cal A}} 
  \def\rd{\mathrm{ d}}

\def\CP{{\cal P}}  

\def\CO{{\cal O}}

\def\bt{\bigtriangledown}

\def\o{\omega}
\def\O{\Omega}

\makeatletter
\def\bard{\protect\@bard}
\def\@bard{%
\relax
\bgroup
\def\@tempa{\hbox{\raise.73\ht0
\hbox to0pt{\kern.4\wd0\vrule width.7\wd0
height.1pt depth.1pt\hss}\box0}}%
\mathchoice{\setbox0\hbox{$\displaystyle\mathrm{d}$}\@tempa}%
{\setbox0\hbox{$\textstyle \mathrm{d}$}\@tempa}%
{\setbox0\hbox{$\scriptstyle \mathrm{d}$}\@tempa}%
{\setbox0\hbox{$\scriptscriptstyle \mathrm{d}$}\@tempa}%
\egroup
}
\makeatother

\makeatletter
\def\barp{\protect\@barp}
\def\@barp{%
\relax
\bgroup
\def\@tempa{\hbox{\raise.73\ht0
\hbox to0pt{\kern.4\wd0\vrule width.7\wd0
height.1pt depth.1pt\hss}\box0}}%
\mathchoice{\setbox0\hbox{$\displaystyle\partial$}\@tempa}%
{\setbox0\hbox{$\textstyle \partial$}\@tempa}%
{\setbox0\hbox{$\scriptstyle \partial$}\@tempa}%
{\setbox0\hbox{$\scriptscriptstyle \partial$}\@tempa}%
\egroup
}
\makeatother

\makeatletter
\def\barb{\protect\@barb}
\def\@barb{%
\relax
\bgroup
\def\@tempa{\hbox{\raise.73\ht0
\hbox to0pt{\kern-.1\wd0\vrule width.7\wd0
height.1pt depth.0pt\hss}\box0}}%
\mathchoice{\setbox0\hbox{$\displaystyle\mathrm{b}$}\@tempa}%
{\setbox0\hbox{$\textstyle \mathrm{b}$}\@tempa}%
{\setbox0\hbox{$\scriptstyle \mathrm{b}$}\@tempa}%
{\setbox0\hbox{$\scriptscriptstyle \mathrm{b}$}\@tempa}%
\egroup
}
\makeatother

   \def\a{\alpha}
   \def\b{\beta}

   \def\de{\delta}
   
\def\l{\lambda}
   \def\e{\epsilon}
   \def\i{\imath}

\def\gr{\mathrm{g}} 
\def\hr{\mathrm{h}}
\def\Gr{\mathrm{G}} 
\def\Gn{\Gr_0}
\def\G1{\Gr_1}
\def\Hr{\mathrm{H}}
\def\GG{\!\Gr/\Gr}
\def\WZW{\!\mathrm{WZW}}
\def\gauge{\!\mathrm{gauge}}
\def\WZ{\!\mathrm{WZ}}
\def\Ad{\mathrm{Ad}}
\def\Adg{\Ad_\gr}
\def\w{\stackrel{\wedge}{,}}
\def\kin{\!\mathrm{kin}}
\def\top{\!\mathrm{top}}\def\DSM{\!\mathrm{DSM}}
\def\HPSM{\!\mathrm{HPSM}}

\def\Skin{S_{\kin}}
\def\Stop{S_{\top}}
\def\SDSM{S_{\DSM}}
\def\we{\wedge}
\def\SGG{S_{\GG}}
\def\dgg{\rd \gr \gr^{-1}}

\def\graph{\mathrm{graph}}
\def\pX{\partial X}
\def\ca{{\tt a}}
\def\tg{{\tt g}}
\def\mg{\tg} 

\newcommand{\simto}{\ensuremath{\to\mbox{\hspace{-1.1em}\raisebox{0.3em}
{$\scriptstyle\sim$}\hspace{0.5em}}}}

\def\LH{\cL_{\mathrm{Ham}}}

\begin{document}
  
\hfill FSU-TPI-08/04\\

\begin{center}
{ \bf \Large  Dirac Sigma Models}
\end{center}

\smallskip

\begin{center}\sl\large
Alexei
Kotov$^{a,c}$\footnote{e-mail address: {\tt kotov@mpim-bonn.mpg.de}},
Peter Schaller$^b$\footnote{e-mail address: {\tt
peter@ca-risc.co.at}}, 
Thomas Strobl$^c$\footnote{e-mail address: {\tt
Thomas.Strobl@uni-jena.de}}
\end{center}


   \begin{center}{\sl $^a$ Max-Planck-Institut f\"ur
   Mathematik, Bonn \\
Vivatsgasse 7, D--53111 Bonn, Germany \\
\vspace{0.2em}
$^b$
Operational Risk and Risk Analysis, 
Bank Austria Creditanstalt, \\
Julius Tandler Platz 3,
A-1090 Vienna,
Austria\\
  \vspace{0.2em}
$^c$ Institut f\"ur
Theoretische Physik, Friedrich-Schiller-Universit\"at Jena, \\
Max-Wienpl. 1, D--07743 Jena, Germany}
\end{center}


\abstract{We introduce a new topological sigma model, whose
fields are bundle maps from the tangent bundle of a 2-dimensional
world-sheet to a Dirac subbundle of an exact Courant algebroid
over a target manifold. It generalizes simultaneously the
 (twisted) Poisson sigma model 
 as well as the $\,\GG$-$\WZW$ model.
 The  equations of motion are satisfied, iff
 the corresponding classical field is a  Lie algebroid morphism.   
 The Dirac Sigma Model
 has an inherently topological part as well as
  a kinetic term which uses a metric on worldsheet and
 target. The latter contribution serves as a kind of regulator for the theory,
 while at least classically
 the gauge invariant content turns out to be independent of any
 additional structure. In the (twisted) Poisson case one may drop the
 kinetic term altogether, obtaining the WZ-Poisson sigma model; in
 general, however, it is compulsory for establishing the morphism property.}

\setcounter{tocdepth}{1}
\tableofcontents

\smallskip

\section{Introduction}

In this paper we introduce a new kind of two-dimensional 
topological sigma model which generalizes  simultaneously the Poisson 
Sigma Model (PSM)
\cite{PSM1,Ikeda,Ctirad}
and the G/G WZW model 
 \cite{GawedzkiGG1,GawedzkiGG2}
 and which  corresponds to general Dirac structures
 \cite{CourantPhD,Diracstructures1}
(in exact Courant algebroids). Dirac structures include Poisson and 
presymplectic structures as particular cases. 
They are $\dim M$-dimensional subbundles $D$ of 
$E := T^*M \oplus TM$ satisfying some particular properties recalled in the 
body of the paper below. If one regards the graph of a contravariant 2-tensor 
$\CP \in \Gamma(TM^{\otimes 2})$ viewed as a map from $T^*M$ to $TM$, then 
$D = \graph \, \CP$ turns out to be a Dirac structure if and only if $\CP$ is a 
Poisson bivector (i.e.~$\{ f, g \} := \CP(\md f, \md g)$, $f,g \in 
C^\infty(M)$,  defines a Poisson  bracket on $M$). Likewise, $D := 
 \graph \, \o$, where $\o$ is a
 covariant 2-tensor viewed as a map $TM \to T^*M$, 
is a Dirac structure, iff $\o$ is a closed 2-form. 

More generally, the above construction is twisted by a closed 3-form
$H$, and in addition  
not any Dirac structure can be written as a graph from $T^*M$ to $TM$ 
or vice versa. This is already true for a Dirac structure that can be
defined canonically on any semi-simple Lie group $\Gr$ and which turns
out to govern the $\,\GG$-$\WZW$ model. Only after cutting out some
regions in the target $\Gr$ of this $\sigma$-model, the Dirac
structure $D$ is the graph of a bivector and the $\,\GG$ model
can be cast into
the form of a (twisted) PSM \cite{G/G,GawedzkiG/G}.  The new 
topological sigma model we are suggesting, the Dirac Sigma Model (DSM), 
works for an arbitrary Dirac structure. 

We remark in parenthesis that also generalized complex structures, 
which lately received increased attention in string theory, fit into the 
framework of Dirac structures. In this case one regards the 
complexification  of $E$ and, as an additional condition, the Dirac
structure $D$ called a ``generalized complex structure''
needs to have trivial intersection with its complex 
conjugate. 
The focus of this text  is on real $E$, but we
intend to present an adaptation separately  (for 
related work cf.~also \cite{Zucchini,Maxim,Bergamin}).


As is well-known, the quantization of the PSM yields the quantization of 
Poisson manifolds \cite{Kontsevich,CF1} (cf.~also \cite{PSM2}).
In particular, the perturbative treatment yields the 
Kontsevich formula. The reduced phase space of the PSM on a strip carries 
the structure of a symplectic groupoid integrating the chosen 
Poisson Lie algebroid \cite{CFHam}. One may expect to obtain similar relations 
for the more general DSM. Also, several two-dimensional field theories
of physical interest were cast into the form of particular PSMs
\cite{PSM1,PartI,Habil1,Kummerreview}
and thus new efficient tools for their analysis were accessible. 
The more general DSMs should permit to enlarge this class of
physics models.




The definition of the DSM requires some auxiliary structures. 
In particular one needs a metric $\tg$ and $h$ on the target manifold
$M$ and  
on the base or worldsheet manifold $\S$, respectively.
The action of the DSM consists
of two parts, $\SDSM =\Stop + \Skin$, where only the ``kinetic'' term
$\Skin$ depends on $\tg$ and $h$. 
If $D = \graph \, \CP$, $\Skin$ may be dropped, at least classically,
in which case one recovers 
the PSM (or its relative, twisted by a closed 3-form, the WZ-Poisson Sigma 
Model). We conjecture that for what concerns the \emph{gauge
invariant} information  captured in the model on the classical level
one may always drop $\Skin$ in $\SDSM$---and for $\Sigma \cong S^1 \times \R$ we
proved this, cf.~Theorem \ref{independence}
below. Still, even classically, it plays an important role,
serving as a kind of regulator for the otherwise less well behaved
topological theory. E.g., in general, it is only the presence of 
$\Skin$ which ensures that the field equations of $\SDSM$ receive the
mathematically appealing interpretation of Lie algebroid morphisms
from $T\S$ to the chosen Dirac structure $D$---in generalization of
an observation for the PSM \cite{BKS}. (We will recall these
notions in the body of the paper, but mention already here that $T\S$
as well as any Dirac structure canonically carry a Lie algebroid
structure).
Without $\Skin$, the solutions of the
Euler Lagrange equations constrain the fields less in general, which
then seems to be balanced by additional gauge symmetries broken by
$\Skin$. These additional symmetries can be
difficult to handle mathematically, since in part they are supported on lower
dimensional regions in the target of the $\s$-model.

The paper is organized as follows:
In Section \ref{sec:GG} we use the
$\,\GG$ model as a starting point for deriving the new sigma model. This
is done by rewriting the $\,\GG$-$\WZW$ model in terms suitable for a
generalization. By construction, the generalization will be such that
the PSM is included, up to $\Skin$, as mentioned above. The role
of the Poisson bivector $\CP$ in the PSM is now taken by an orthogonal
operator $\CO$ on $TM$, which in the Poisson case is related to $\CP$
by a Cayley transform, but which works in the general case. 

In Section \ref{Background} we provide the
mathematical background that is necessary for a correct interpretation
of the structures defining the general sigma model. This turns out to
be the realm of Courant algebroids and Dirac structures. We
recapitulate
definitions and facts known in the mathematics
literature, but also original results, developed to address the needs
of the sigma model, are contained in this section.
The action of the Dirac sigma model
is then recognized as a
particular functional on the space of vector bundle morphisms 
$\phi \colon T\S \to D$, $\SDSM = \SDSM [\phi]$. Specializing this to
the PSM, one reproduces the usual fields $\tilde \phi \colon T\S
\to T^*M$, since precisely in this case $D$ is isomorphic to $T^*M$.

In Section \ref{sec:split} we point out that the definition of the DSM
presented in the preceding sections also depends implicitly on some
further auxiliary structure in addition to the chosen Dirac structure
$D \subset E$,
namely on a ``splitting'' in the exact Courant algebroid $\,0 \to T^*M
\to E \to TM \to 0$. This dependence occurs in $\Stop$, but again at
the end of the day, the ``physics'' will not be effected by it.

In
Section \ref{sec:eom} we derive the field equations of $\SDSM$, which
we present in an inherently covariant way. We also 
prove that $\phi$ solves the field equation, iff it respects the
canonical Lie algebroid structures of $T\S$ and the Dirac structure
$D$,
i.e.~iff  $\phi \colon T\S \to D$ is a Lie algebroid morphism. 
We present one possible covariant (global and frame
independent) form
of the gauge symmetries of $\SDSM$, furthermore,
using the connection on $M$
induced by the auxiliary metric $\tg$. We will, however, postpone the
corresponding proof of the gauge invariance and  further
interpretations to another work \cite{AT1}, where the question
of covariant gauge symmetries will be addressed in a more general framework
of Lie algebroid theories, for which the DSM serves as one possible
example. There we will also relate these
symmetries to the more standard presentations of the symmetries of the
$\, \GG$ model and the (WZ-)PSM.

In Section \ref{sec:Hamilton},
finally,
which in most parts can be read also directly after Section \ref{sec:GG}, 
we determine the Hamiltonian structure of the DSM. In
fact, we will do so even for a somewhat more general sigma model,
where the target subbundle $D \subset E$ is not necessarily
required to be integrable. It turns out that the constraints of this
model are  of the form introduced recently in \cite{AS}, where now
currents $J$ are associated to any section $\psi \in \Gamma(D)$. As a
consequence of the general considerations in \cite{AS}, the constraints
$J_\psi = 0$ are found to be first class, iff $D$ is integrable,
i.e.~iff it is a Dirac structure.


\section{From G/G to Dirac Sigma Models}
\label{sec:GG}

We will use the G/G-WZW-model  \cite{GawedzkiGG1,GawedzkiGG2}
as a guide to the new sigma model that
is attached to any Dirac structure. Given a Lie group $\Gr$ with
quadratic Lie algebra $\g$ and a closed 2-manifold $\S$ equipped with
some metric $h$, which we assume to be of Lorentzian signature for
simplicity, the
(multivalued) action functional of the
WZW-model consists of two parts (cf.~\cite{WZW} for further details):
a kinetic term for
$\Gr$-valued fields $\gr(x)$ on $\S \ni x$ as
well as a Wess-Zumino term $S_{\WZ}$, requiring the (non-unique)
extension of $im \S \subset
\Gr$ to a 3-manifold $N_3 \subset \Gr$ such that $\partial N_3 = im
\S$:
\beq \label{WZW1} S_{\WZW} [\gr] =
\frac{k}{4\pi} \int\limits_\Sigma \langle \partial_+ \gr \gr^{-1},
\partial_- \gr
\gr^{-1} \rangle \: \rd x^- \wedge \rd x^+ + S_{WZ} \\
S_{\WZ} [\gr] =\frac{k}{12\pi}
\int\limits_{N_3} \langle \rd \gr \gr^{-1} \w
(\rd \gr \gr^{-1})^{\wedge 2}\rangle \, ,\label{WZ}
\eeq
where $x^+,x^-$ are lightcone coordinates on $\S\,$
(i.e.~$h=\rho(x^+,x^-) \, \rd x^+ \rd
x^-$ for some locally defined positive function $\rho$), $\langle
\cdot, \cdot \rangle$ denotes the Ad-invariant scalar product on
$\g$, and $k$ is an
integer multiple of $\hbar$
(which implies that the exponent of $\frac{i}{\hbar} S_{\WZW}$,
the integrand in a path integral, is a unique functional of $\gr
\colon \S \to \Gr$). Introducing a connection 1-form $a$ on $\S$ with
values in a Lie subalgebra $\h < \g$, one can lift the obvious rigid
gauge invariance of (\ref{WZW1}) w.r.t.~$\gr \mapsto \Ad_\hr \gr
\equiv \hr \gr \hr^{-1}$, $\hr
\in \Hr < \Gr$, to a local one ($\hr=\hr(x)$ arbitrary, $a \mapsto
\hr \rd \hr^{-1} + \Ad_\hr a$)
by adding to $S_{\WZW}$
\beq S_{\gauge}[\gr,a] = \frac{k}{2\pi} \int_\S \left( \langle
a_+,\partial_- \gr \gr^{-1} \rangle -  \langle
a_-, \gr^{-1}  \partial_+ \gr\rangle +  \langle
a_+,a_- \rangle -\langle
a_+, \gr a_- \gr^{-1} \rangle \right) \, \rd^2 x \, ,
\eeq
where $\rd^2 x \equiv \rd x^- \wedge \rd x^+$.
For the maximal choice $\Hr=\Gr$ this yields the G/G-model:
\beq S_{\GG}[\gr,a] = S_{\WZW}[\gr] + S_{\gauge}[\gr,a] \, . \label{GG1} \eeq

In \cite{G/G} it was shown that on the Gauss decomposable part
$\Gr_{\mathrm{Gauss}}$
of $\Gr$ (for $SU(2)$ this is all of the 3-sphere except for a 2-sphere)
the action (\ref{GG1}) can be replaced equivalently by a
Poisson Sigma Model (PSM) with target $\Gr_{\mathrm{Gauss}}$.
(This was re-derived in a more
covariant form in \cite{GawedzkiG/G}). It is easy to see
 that by
similar manipulations---and in what follows we will demonstrate this by
a slightly different procedure---(\ref{GG1}) can be cast into a WZ-PSM
\cite{Ctirad} on
$\Gr_1 :=\Gr\backslash \Gr_0$, where $\Gr_0=\{ \gr \in \Gr | \ker(1+\Ad_\gr)
\neq \{ 0 \} \}$ (again a 2-sphere for  $SU(2)$). The
question may arise, if there do not exist possibly some other manipulations
that can cast the G/G-model into the form of a WZ-PSM \emph{globally},
with a 3-form $H$ of the same cohomology as the Cartan 3-form (the
integrand of (\ref{WZ})). In fact this is not possible: it may be
shown \cite{AKS} that there is a cohomological obstruction for writing
the Dirac structure which governs the G/G-model and which is disclosed below
(the Cartan-Dirac structure, cf.~Example \ref{ex:Cartan} below)
globally as a graph
of a bivector. Consequently, this calls for a new type of topological
sigma model that can be associated to \emph{any} Dirac structure $D$ (in
an exact Courant algebroid) such that it specializes to the WZ-PSM if
$D$ may be represented as the graph of a bivector and e.g.~to the G/G
model if the target $M$ is chosen to be $\Gr$ and $D$ the Cartan-Dirac
structure.\footnote{We remark in parenthesis that in an Appendix in
\cite{G/G} it was shown that the G/G model can be represented on all
of $\Gr$ as what we would call these days a WZ-PSM, but this was at the
expense of permitting a \emph{distributional} 3-form $H$ (the
support of which was on $\Gr\backslash\Gr_{\mathrm{Gauss}}$). The
above mentioned topological 
obstruction relies on the smooth category.}

Vice versa, the G/G model already provides a possible
realization of the sought-after sigma model
for this particular choice of $M$ and $D$. We will thus use it to
derive the new sigma model within this section. For this purpose it
turns out profitable to rewrite (\ref{GG1}) in the language of
differential forms, so that the dependence of the worldsheet metric
$h$ becomes more transparent; for simplicity we put $k=4\pi$ in what
follows (corresponding to a particular choice of $\hbar$). We then
find
\beq
S_{\GG} &=& \frac{1}{2} \int\limits_\Sigma \langle \rd\gr \gr^{-1}\w * \rd\gr
\gr^{-1} \rangle + S_{\WZ} + \int\limits_\Sigma  \langle a\w*a\rangle -
\langle a\w  \Ad_\gr (*-1)a \rangle 
\\{}&&  + \int\limits_\Sigma\langle a\w (*-1)
\rd\gr \gr^{-1} \rangle -
\langle \Ad_\gr a\w (*+1) \rd\gr \gr^{-1}
\rangle   \, .
 \label{GG2} \eeq
Here our conventions for the Hodge dual operator $*$, which is the
only place where $h$ enters, is such that  $* \rd x^\pm = \pm \rd
x^\pm$. Now we split $S_{\GG}$ into terms containing $*$ and those
which do not. One finds that the first type of terms combines into a
total square and that 
\beq
S_{\GG}[\gr,a] &=& S_{\kin} + S_{\top}  
\label{GG3} \\ S_{\kin} &=& \frac{1}{2} \int_\S || \rd \gr \gr^{-1} +
(1-\Ad_\gr) a ||^2
\label{GGSkin}
\\  S_{\top} &=& \int_\S
\langle -(1+\Ad_\gr) a \w \rd\gr \gr^{-1} \rangle +  \langle a \w
\Ad_\gr a\rangle
+ S_{\WZ}\label{GGStop}
\eeq
where for a Lie algebra valued
1-form $\beta$ 
we use the notation
$|| \beta ||^2 \equiv \langle \beta \w * \beta \rangle$.

\vskip4mm

Before generalizing this form of the action, we show that G/G can be
cast into the form of a WZ-PSM (or HPSM for a given choice of $H$)
on $\Gr_1$. For this purpose we briefly recall the action functional
of the WZ-PSM \cite{Ctirad} (cf.~also \cite{Park}): Given a closed
3-form $H$ and a bivector field $\CP = \frac{1}{2} \CP^{ij}(X) \,
\partial_i \wedge \partial_j$ on a target manifold $M$ one
considers
\beq
S_{\HPSM}[X,A] = \int_\S A_i \wedge \rd X^i + \2 \CP^{ij} A_i \wedge A_j +
\int_{N_3} H  \, , \label{HPSM}
\eeq
where  $X \colon \S \to M$ and $A \in \Gamma(T^*\S \otimes X^*T^*M)$
and the last term is again a WZ-term (i.e.~$N_3 \subset M$ is 
chosen such that its boundary agrees with the image of $X$ and usual
remarks about multi-valuedness of the functional can be made); for the
case that $H = \rd B$ the latter contribution can be replaced by (the
single-valued) $\int_\S \2 B_{ij} \, \rd X^i \wedge \rd X^j$. This
kind of theory is topological (has a finite dimensional moduli space
of classical solutions modulo gauge transformations) iff the couple
$(\CP,H)$ satisfies a generalization of the Jacobi-identity, namely
\beq \label{HJacobi}\CP^{il} \partial_l \CP^{jk} + {\rm cycl}(ijk)
= H_{i'j'k'} \,  \CP^{i'i} \CP^{j'j} \CP^{k'k} \, ;
\eeq
$(\CP,H)$ then defines a WZ-Poisson structure on $M$ (also called
``twisted Poisson'' or ``$H$-Poisson'' or ``Poisson with background''
in the literature). 
 
We now want to
show that when restricting $\gr$ to maps  $\gr
\colon \S \to \Gr_1$, the action $S_{\GG}$ can be replaced by
(\ref{HPSM}) for a particular choice of  $\CP$ and $H$, at least for
what concerns the classical field equations. Let us consider the
variation of the two contributions to $\SGG$ in (\ref{GG1}) with
respect to $a$ separately:
\beq - \Adg \frac{\delta \Skin}{\delta a} = (1-\Adg) * \left[ \dgg +
(1-\Adg) a \right] \label{eom1a} \\ - \Adg \frac{\delta \Stop}{\delta a} =
(1+\Adg)  \left[ \dgg + (1-\Adg) a \right] \, . \label{eom1b}
\eeq
Decomposing the term in the square bracket into its $\pm 1$
eigenvalues of $*$ (use projectors $\2 (1 \pm *)$ or, equivalently,
consider the $\rd x^+$ and $\rd x^-$ components of (\ref{eom1b})), it
is easy to see
that $\delta \SGG /\delta a = 0$ yields
\beq \dgg + (1-\Adg) a  = 0 \, . \label{eom1} \eeq
On the other hand, as obvious from (\ref{eom1b}), this equation is also
obtained from $\Stop$ on $G_1$. Since $\Skin$ is quadratic in the
left-hand side of (\ref{eom1}), it gives no contribution to the
variation of $\SGG$ w.r.t.~$\gr$, which proves the desired
equivalence.

For completeness we remark that the second field equation
is nothing but the zero curvature condition\footnote{Written in 
a matrix representation. More generally, $F \equiv  \md a + \2 [a \w a]$.} 
\beq F \equiv \md a + a \we a = 0  \label{eom2} \, , \eeq
and that this equation results from variation of $\Stop$ w.r.t.~$\gr$
even on all of $\Gr$. Note however that $\Stop$ will have solutions
mapping into $\Gn = \Gr \backslash \G1$ violating (\ref{eom1}). 

It thus remains to cast $\Stop$ into the form (\ref{HPSM}). On $\G1$
this is done most easily by introducing $A := -(1 + \Ad_\gr) a$, where
the matrix components of this 1-form corresponds to a right-invariant
basis of $T^*\Gr$ (note also that sections of  $T^*\Gr$ and  $T\Gr$,
1-forms and vector fields on the group, can be identified by means of
the Killing metric); then the first term in (\ref{GGStop}) already
takes the form of the first term in (\ref{HPSM}). The WZ-terms can be
identified without any manipulations. It remains to
calculate the bivector upon comparison of the respective second
terms. A very simple calculation then yields
\beq
 \CP  = \frac{1-\Adg}{1+\Adg} \; , \quad H = \frac{1}{3}
 \langle \rd \gr \gr^{-1} \w
(\rd \gr \gr^{-1})^{\wedge 2}\rangle \, , \label{GGHPoisson}
\eeq
where $\CP$ refers to a right-invariant basis on $\Gr$ again and $H$
is the Cartan 3-form. In \cite{3Poisson} it was shown that the
WZ-Poisson structures \cite{Ctirad} are particular Dirac structures;
and the utility of this reformulation was stressed due to  the
simplification of checking (\ref{HJacobi}) for the above example. We
want to use the opportunity to stress the usefulness of sigma models
in this context (cf.~also \cite{talkESI01}): Using the well-known fact
that the G/G model is
topological (in the sense defined above) and that one can cast it into
the form (\ref{HPSM}) is already sufficient to establish
(\ref{HJacobi}) for (\ref{GGHPoisson}); even more, the above
consideration is a possible route for finding this example of
a WZ-Poisson or Dirac structure. The above bivector also plays a
role in the context of D-branes in the WZW-model.

\vskip4mm

We now come to the generalization of the G/G model written in the form
(\ref{GG3}). For this purpose we first rewrite (\ref{HPSM}) into a
form more suitable to the language of Dirac structures. It is
described as the graph of the bivector $\CP$ in the bundle $E=T^*M
\oplus TM$, i.e.~as pairs $(\alpha, \CP(\alpha,\cdot))$ for any
$\alpha \in T^*M$. With the 1-forms $A$ taking values in $T^*M$ we
thus may introduce the dependent 1-form $V=\CP(A,\cdot)$ taking values
in $TM$. Together they may be viewed as a 1-form $\CA = A \oplus V$ on
$\S$ taking values in the subbundle $D = \graph(\CP) \subset E$. 
Then (\ref{HPSM}) can be rewritten as
\beq
\Stop[X,\CA] = \int_\S A_i \wedge \rd X^i - \2  A_i \wedge V^i +
\int_{N_3} H  \, , \label{Stop}
\eeq
Comparison with (\ref{GGStop}) shows that in the G/G-model $A_\a =
-\left[(1+\Adg)a\right]_\a$, where $\a$ is an index referring to a
right-invariant basis on $\Gr$. Correspondingly, we can read off by
comparison of  (\ref{GGStop}) with (\ref{Stop}) that then $V^\a = 
-\left[(1-\Adg)a\right]^\a$ (showing equality is a simple exercise
where one uses that $\Adg$ is an orthogonal operator w.r.t.~the
Killing metric). Note that in the formulation (\ref{Stop}) no metric
on $M$ appears anymore; the Killing metric is used only in the above
identification. 

The G/G-model also contains a second part, which uses a metric $h$ on
$\S$ as well as a metric $\mg$ on $M$. From the above identification it
is easy to generalize it:
\beq \Skin[X,\CA] = \frac{\a}{2} \int_\S || \rd X - V ||^2 \, ,
\label{Skin} \eeq
where for any $TM$-valued 1-form $f = f^i \partial_i =
f^i_\mu \rd x^\mu \otimes \partial_i$ on $\S$ we use 
\beq \label{eq:norm}
|| f ||^2 := \mg(f  \!    \w \!
 * f) \equiv  \mg_{ij} h^{\mu \nu}
f^i_\mu f^j_\nu \: \mathrm{vol}_\S \, , \quad  \mathrm{vol}_\S \equiv
\sqrt{\det( h_{\mu \nu})} \rd ^2 x \, , \eeq
and where $\a$ is some
coupling constant. 
For the action functional of the Dirac Sigma Model (DSM) we thus postulate
$ \SDSM [X,\CA]:=  \Skin + \Stop $, i.e.
\beq \SDSM [X,A \oplus V] = \frac{\a}{2} \int_\S || \rd X - V ||^2 +
\int_\S A_i \wedge \rd X^i - \2  A_i \wedge V^i +
\int_{N_3} H  
\, . \label{DSM} \eeq
As already mentioned above, the 1-forms $\CA \equiv A \oplus V$
take value in $X^*D$,
where $D$ is a Dirac structure; this will be made more precise and
explicit below. For Lorentzian signature metrics $h$ on $\S$, $\a$
should be real and (preferably) non-vanishing; for Euclidean signatures
of $h$ we need an imaginary unit as a relative factor between the
kinetic and the topological term. Although possibly unconventional, we
will include it in the coupling constant $\alpha$ in front of the
kinetic term, so that we are able to cover all possible signatures
in one and the same action functional. If $\tg$ has an indefinite
signature, on the other hand, we in addition need to restrict to a
neighborhood of the original value $\a = 1$ (or $\a = i$ for Euclidean
$h$); the condition we want to be fulfilled is the invertibility of
the operator (\ref{eq:op}) below (cf.~also the text following
Corollary \ref{corop}).

\vskip4mm

The metrics $h$ and $\mg$ on $\S$ and $M$, respectively, are of
auxiliary nature. First of all, it is easy to see, that $\Skin$ gives
\emph{no} contribution to the field equations for what concerns the
WZ-PSM (\ref{HPSM}); also the gauge symmetries are modified only slightly
by some on-shell vanishing, and thus physically irrelevant
contribution (
both statements will be proven explicitly in subsequent sections). 
Let us consider the other extreme case of a Dirac structure provided by 
the subbundle $D=TM$ to $E=T^*M\oplus TM$ (for $H$ being zero):
In this case $A \equiv 0$
and $V=V^i \partial_i$ is an unconstrained 1-form field. Obviously in
this case $\Stop \equiv 0$ and one obtains \emph{no} field equations from
this action alone. On the other hand, the field equations from $\Skin$ 
are computed easily as
\beq \rd X^i = V^i \, . \label{TMmorph} \eeq
First we note that this equation does not depend neither on $h$ nor on
$\mg$; these two structures
 are of auxiliary nature for obtaining a nontrivial field
equation in this case, a fact that will be proven also for the general
case in the subsequent section (cf.~Theorem \ref{theo1}
below). Secondly, we observe that at the end of the day even in this
case the two theories $\Stop[X,\CA] \equiv 0$ and $\SDSM[X,\CA] \equiv
\Skin[X,V]$ are still not so different as one may expect at first sight:  
The \emph{moduli space} of classical solutions is the same for both
theories. The lack of field equations in the first case is compensated
precisely in the correct way by additional gauge symmetries, that are
absent for $\Skin[X,V]$. If we permit as gauge symmetries those that are
in the connected component of unity, we find the homotopy classes
$[X]$ of $X \colon \S \to M$ as the only physically relevant
information. (Gauge identification of different homotopy classes might
be considered as large gauge transformations, as
both action functionals remain unchanged in value; then the
moduli space would be just a point in each case). 

One may speculate that this mechanism of equivalent moduli spaces
occurs also in the general situation. We leave this as a conjecture
for a general choice of $\S$, proving it in the case of
$\S = S^1 \times \R$, where we will  establish equivalent
Hamiltonian structures (cf.~section \ref{sec:Hamilton} below).
We remark, however, that the equivalence may require some slightly
generalized notion of gauge symmetries similar to the $\l$-symmetry
discussed in \cite{AYM}; this comes transparent already from the G/G
example, where the additional classical solutions in the kernel of
$1+\Adg$ found above need to be gauge identified by \emph{additional}
gauge symmetries of $\Stop$ that are concentrated at the same region in $\Gr$. 

Note that this complication disappears when $\Skin$ is added to
$\Stop$. Likewise, the field equations (\ref{TMmorph}) have a nice
mathematical interpretation; they are equivalent to the statement that
the fields $(X,\CA)$ are in one-to-one correspondence with morphisms
from $T\S$ to $D$, both regarded as Lie algebroids (cf.~Theorem \ref{theo1}
below). So, the addition of $\Skin$ (with non-vanishing $\a$)
serves as a kind of regulator for the
theory, making it mathematically more transparent and  more
tractable--while simultaneously the ``physics'' (moduli space of
solutions) seems to remain unchanged in both cases, $\a = 0$ and $\a
\neq 0$. 

\vskip4mm

Having an auxiliary metric $\mg$ on $M$ at disposal, one may use it
profitably to reformulate $\Stop$. In particular, it will turn out
that one may use it to parameterize the Dirac structure
\emph{globally} in
terms of an orthogonal operator $\CO \colon TM \to TM$
(cf.~Proposition \ref{graph} below), generalizing the operator $\Adg$
on $M=\Gr$ in the G/G model above; this then permits one to use
unrestricted fields for the action functional again, such as $\gr$ and
$a$ in the G/G model and $X$ and $A$ in the WZ-PSM. 

Essentially, this works as follows:
By means of $\mg$ we may identify $T^*M$ with $TM$, so $E \cong TM
\oplus TM$ and the parts $A$ and $V$ of $\CA = A \oplus V$ may be
viewed both as (1-form valued) vector (or covector) fields on $M$
(corresponding to the index position, where indices are raised and
lowered by means of $\mg$). 
Introducing the involution $\tau \colon E \to E$ that exchanges both
copies of $TM$, $\tau (\a \oplus v) = v \oplus \a$, let us consider
its eigenvalue subbundles $E_\pm = \{ (v \oplus \pm v) \, , \; v \in
TM \}$, both of which can be identified with $TM$ by projection to the
first factor $T^*M \cong TM$.
It turns out (cf.~Proposition \ref{graph} below) that
any Dirac structure $D \subset E$ can be regarded as the graph of a
map from $E_+ \to E_-$, which, by the above identifications, corresponds
to a (point-wise) map $\CO$ from $TM$ to itself. Let us denote the 
$E_+$ and $E_-$ decomposition of an element of $E$ as $(v_1 ; v_2)$, 
where elements $v_i$ may be regarded as vectors on $M$. Then any Dirac
structure can be written as $D = \{
(v ; \CO v) \in E , v \in TM\}$, where $\CO$ is point-wise
orthogonal w.r.t.~the metric $\mg$. Obviously, 
$(v ; \CO v) = (1+\CO)v \oplus (1-\CO)v \in TM \oplus TM \cong T^*M
\oplus TM =E$; thus e.g.~the graph $D = \graph(\CP)$ of a bivector
field $\CP$, $D = \{ \a \oplus \CP(\a,\cdot) , \a \in T^*M \} \subset
E$ corresponds to the orthogonal operator
\beq \CO = \frac{1-\CP}{1+\CP} \quad \Leftrightarrow \quad
 \CP = \frac{1-\CO}{1+\CO} \, . \label{Cayley}
\eeq 
Note that in a slight abuse of notation we did not distinguish between
the bivector field $\CP = \frac{1}{2}\CP^{ij} \partial_i \wedge
\partial_i \in \Gamma(\Lambda^2 TM)$, the canonically induced map from
$T^*M \to TM$, $\a \mapsto  \CP(\a,\cdot)$,  and the corresponding
operator on $TM$ using the isomorphism induced by $\mg$: $TM \ni v
\mapsto \CP( \mg(v,\cdot),\cdot) \in TM$; in particular this implies that in
an explicit matrix calculation using some local basis $\partial_i$ in
$TM$, with $\CO = \CO^i_j \partial_i \otimes \rd X^i$, 
the matrix denoted by $\CP$ in (\ref{Cayley}) is $\CP_i{}^j
\equiv \mg_{ik} \CP^{kj}$. 

Obviously the Dirac structure of the G/G model corresponds to the
choice $\Ad_\gr$ for $\CO$ above and the first formula
(\ref{GGHPoisson}) is the specialization of (\ref{Cayley}) to this
particular case. The transformation (\ref{Cayley}) is a Cayley map.
Although any antisymmetric matrix $\CP$ yields an orthogonal matrix
$\CO$ by this transformation, the reverse is not true. This is the
advantage of using $\CO$ over $\CP$, as it works for any Dirac
structure.\footnote{This observation is due to the collaboration of
A.K.~and T.S.~with A.~Alekseev and elaborated further in \cite{AKS}.
We are also grateful to A.~Weinstein for pointing out to us that the
description of Dirac structures by means of sections of $O(TM)$ was
used already in the original work \cite{CourantPhD}.}
Certainly such as the bivector of  WZ-Poisson structure has to satisfy an
integrability condition, namely equation (\ref{HJacobi}),
which for $H=0$  states that $\CP$ defines a Poisson structure.
There is a likewise
condition to be satisfied by $\CO^i_j(X)$ so that, more generally,
$\CO$ describes a Dirac structure. $\CO$ corresponds to a Dirac
structure iff $U := 1 -\cO$ satisfies
(cf.~Proposition \ref{prop:O} below):
\beq {U^{\tilde i}}_i {U^{\tilde j}}_{j;{\tilde i}} (1-U)_{{\tilde
j}k} + cycl(i,j,k)=
         \2 H_{\tilde i\tilde j\tilde k}{U^{\tilde i}}_i {U^{\tilde
         j}}_j
         {U^{\tilde k}}_k \, . \label{U}
\eeq
Here the semicolon denotes the covariant derivative with respect to
the
Levi Civita connection of $\tg$. 
Locally it may be replaced by an ordinary partial derivative, if the auxiliary
metric is chosen to be flat on some coordinate patch.

Having characterized $D$ by $\CO \in \Gamma(O(TM))$, we may
parameterize $\CA \in \Omega(\S,X^* D)$ more explicitly by $a=a^i
\partial_i \in
\Omega(\S,X^* TM)$ according to
$\CA = -(1+\CO) a \oplus -(1-\CO) a$.
Then the total action (\ref{DSM})
can be rewritten in the form 
\beq\nonumber
\SDSM[X,a]&=& 
 \frac{\a}{2}\int\limits_\Sigma ||\rd X + (1-\CO)a||^2  
+ \int\limits_\Sigma \mg( \rd X
 \w(1+\CO)a ) + \mg(  a  \w \CO a) +   \int\limits_{N_3} H
 \\ \label{practDSM} &\equiv& \frac{\a}{2}\int\limits_\Sigma (\rd X^i + a^i -
\CO^i_k a^k) \wedge \ast  (\rd X^j + a^j -
\CO^j_m a^m)\, \mg_{ij} \\ \nonumber
 && +  \int\limits_\Sigma  \rd X^i \wedge  a^j
\, (\mg_{ij} + \CO_{ij}) +  \CO_{ij} a^i \wedge a^j +
\int\limits_{N_3} H 
\eeq
where now $X^i$ and $a^i$, local 0-forms and 1-forms on $\S$,
respectively,  can be varied without any constraints and indices are
lowered and raised by means of $\mg_{ij}(X)$ and $\mg^{ij}(X)$,
respectively. 
We  stress again that the $\mg$-dependence of the last line is
ostensible only, whereas the $\a$-dependent terms depend on it
inherently. 

The above presentation of
$\CA$ was suggested by the $\, \GG$ model. In the rest of the paper
we will however rather use the slightly more elegant parameterization 
$\CA = (1+\CO) \ca \oplus (1-\CO) \ca$,
resulting from  $\ca := -a$. In these variables  the action
(\ref{practDSM}) takes the form 
\beq
\label{eq:DSMca}
\SDSM[X,\ca]= 
 \frac{\a}{2}\int\limits_\Sigma ||f||^2  
+ \int\limits_\Sigma \mg(
 (1+\CO)\ca \w \rd X ) + \mg(  \ca \w \CO \ca) +   \int\limits_{N_3} H
\eeq
where we used (\ref{eq:norm}) with 
\beq f \equiv \rd X - V \equiv  \rd X - (1-\CO)\ca \, . \eeq

In the following sections we will tie the above formulas to a more
mathematical framework and, among others,
analyze the field equations from this perspective.
Readers more interested in applications for physics may also be
content with consulting only the main results from the following
sections, in particular Theorem \ref{theo1} and Proposition
\ref{prop:gaugesymmetry}, and then turn directly to 
the Hamiltonian analysis of the action
in Section \ref{sec:Hamilton}.

Noteworthy is maybe also the generalization 
Eq.~(\ref{eq:newkin}) of the kinetic term introduced in Section \ref{sec:split}
below. The modification of the old kinetic term
uses a 2-form $C$ on the target and is
independent of any metric. Such a generalized kinetic term is
suggested by the more mathematical considerations to follow, but
will not be pursued any further within the present paper.

\vskip4mm


We close this section with a continuative remark: 
As mentioned above, for non-vanishing parameter $\alpha$, the classical
theory will turn out to not depend on this parameter. It is tempting to
believe that this property can be verified also on the quantum level,
a change in $\a$ corresponding to the addition of a BRS-exact term.
In this context it may be interesting to
regard the limit $\alpha \to \infty$, yielding localization to $f=0$.
In fact, one may expect localization of the path integral to all
equations of motion, cf.~\cite{Witten2dYM,GerasimovGG}.




\section{
Dirac structures}\label{Background}
The purpose of this section is to provide readers with the mathematical 
background for the structures used to define the Dirac sigma model.
 We review some basic facts about Dirac structures, being
maximally isotropic (Lagrangian) subbundles in an exact Courant algebroid, the 
restriction of the Courant bracket to which is closed. We describe an explicit
isomorphism between the variety of all Lagrangian subbundles and the group
of point-wise acting operators in the tangent bundles, orthogonal
with
respect to a fixed Riemann metric. We derive an obstruction for such
operators to represent a Dirac structure, cf.~Proposition \ref{prop:O}
below.

 A {\sl Courant algebroid} 
\cite{Diracstructures1,CourantPhD} is a vector bundle $E$ equipped with a
non-degenerate symmetric bilinear form $\langle , \rangle$,
a bilinear operation $\circ$ on $\Gamma (E)$ (sometimes also denoted
as a bracket $[ \cdot , \cdot]$), and
a bundle map $\rho :E\to TM$ satisfying the following properties:

\begin{enumerate}
\item   The left Jacobi condition 
$e_1\circ (e_2\circ e_3)=(e_1\circ e_2)\circ e_3+e_2\circ (e_1\circ e_2)$
\item Representation $\rho (e_1\circ e_2)=[\rho (e_1),\rho (e_2)]$
\item Leibniz rule $e_1\circ fe_2=fe_1\circ e_2+L_{\rho (e_1)}(f) e_2 $
\item $e\circ e=\frac{1}{2}\cD \langle e,e\rangle$
\item Ad-invariance $\rho (e_1)\langle e_2,e_3\rangle =\langle e_1\circ
e_2,
e_3\rangle
+\langle e_2,e_1\circ e_3 \rangle $
\end{enumerate}
where $\cD$ is defined as 
$\cD: C^{\infty}(M)\stackrel{\rd}{\to} \Omega^1 (M)\stackrel{\rho^*}{\to}
 E^* \simeq E$. Properties 2 and 3 can be shown to follow  from
 the other three properties, which thus may serve as axioms
 (cf.~e.g.~\cite{YKS03}).
A Courant algebroid is called {\sl exact} \cite{Severalett}, 
if the following sequence
is exact:
\beq
0\rt T^* M \stackrel{\rho^*}{\to}E\stackrel{\rho}{\to}TM\to 0
\label{exact}
\eeq
A {\sl Dirac structure} $D$ in an exact Courant algebroid is a maximally
isotropic (or Lagrangian) subbundle with respect to the scalar product,
which is closed under the product. A Dirac structure 
is always a particular Lie algebroid: By definition a Lie algebroid is a 
vector bundle $F \to M$ together with a bundle map $\rho \colon F \to TM$ 
and an antisymmetric product (bracket)
between its sections satisfying the first three properties in the list 
above (where again the second property can be derived from the other two). 
In particular, the product, often also denoted as a bracket, $e_1 \circ
e_2 :\equiv [e_1, e_2]$, defines an infinite dimensional Lie algebra
structure on $\Gamma(F)$. 
Due to the isotropy of $D$, the induced product (bracket)
becomes skew-symmetric and 
obviously $D$ is a Lie algebroid. 
 


{}From now on will only consider exact Courant algebroids (\ref{exact}).
Let us choose a "connection" on $E$, i.e. an isotropic splitting
 $\sigma : TM\to E$, $\rho \circ \sigma = \mathrm{id}$. The difference
\beq
\sigma (X)\circ \sigma (Y)-\sigma ([X,Y])=\rho^* H (X,Y) \label{Hdef}
\eeq
is a pull-back of a $C^\infty(M)-$linear, completely skew-symmetric tensor
$H\in \Omega^3(M)$, given by $H(X,Y,Z)= \langle \sigma(X) \circ 
\sigma(Y) , \sigma(Z) \rangle$. From the above axioms one may  deduce
the "Bianchi identity": $\rd H =0$.

Once a connection is chosen, any other one differs by the graph
of a $2-$form $B$. Its curvature is equal to
$H +\rd B $. Therefore the cohomology class $[H ]\in H^3(M)$
is completely determined by the Courant algebroid \cite{Severalett}.

Choosing a splitting with the curvature 3-form $H$, it is possible to
identify the
corresponding exact Courant algebroid with $T^*M\oplus TM$ and the
scalar product with the natural one:
\beq\label{prod1}
\langle \xi_1 +\theta_1 ,\xi_2 +\theta_2 \rangle =
\theta_1 (\xi_2 )+\theta_2 (\xi_1 )\; ,
\eeq
where $\xi_i\in \Gamma (TM)$, $\theta_i \in \Omega^1 (M)$.
Finally, the multiplication law can be shown to take the form: 
\beq\label{mult1}
(\xi_1 +\theta_1)\circ (\xi_2 +\theta_2 ) =
[\xi_1 ,\xi_2 ] + L_{\xi_1}\theta_2 -\i_{\xi_2}\md\theta_1 +
H(\xi_1 ,\xi_2 , \cdot) \; ,
\eeq
where $L_\xi$
and $i_\xi$ denote Lie derivative along a vector field $\xi$ and contraction with $\xi$,
respectively.

Let $E$ be an exact Courant algebroid with a chosen
splitting $E=T^*M\oplus TM$ and a vanishing 3-form curvature
(this implies that the chacteristic 3-class of $E$ is trivial). Then
we have:
\begin{example} 
\label{ex1} Let $D$ be a graph of a 
Poisson bivector field $\cP\in \Gamma (\Lambda^2 TM)$ considered as a
skew-symmetric map
from $T^*M$ to $TM$, then $D= \{ \theta \oplus \cP(\theta) \}$ is a Dirac
subbundle and  the projection from $D$  to $T^*M$ is bijective.
Any Dirac subbundle
 of $E$ with  bijective projection to
 $T^*M$ is a graph of a Poisson  bivector field.
\end{example}
\begin{example} \label{ex2} Let $D$ be a graph of a closed 
$2-$form $\omega\in \Omega^2 (M)$ considered as a skew-symmetric map
from $TM$ to $T^*M$, then $D= \{ \omega(v) \oplus v \}$ is a Dirac
subbundle and  the projection of $D$  to $TM$ is bijective.
Any Dirac subbundle
 of $E$ for which the projection to
  $TM$ is bijective is a graph of a closed $2-$form.
\end{example}
If $E$ has a non-trivial characteristic class $[H]$, one needs to replace 
Poisson (presymplectic) structure by a corresponding WZ (or twisted)
one, cf.~\cite{Ctirad,3Poisson}.

Now we describe
all Lagrangian subbundles of an exact Courant algebroid
$E$, which are not necessarily 
projectable, neither to $T^*M$ (for any splitting $\sigma$) nor to $TM$
(cf.~also \cite{CourantPhD}).
Let us choose an arbitrary 
Riemannian metric $\mg$ on $M$, which can be thought of as a non-degenerate
 symmetric map from $TM$ to $T^*M$. The inverse of 
$\mg$ is acting from $T^*M$ to $TM$. We denote these actions as
$\xi \mto \xi^*$ and $\theta \mto \theta^*$ for a vector field $\xi$ and a 1-form $\theta$, 
respectively. In some local coordinate chart it can be written as follows:
\beq
\label{eq:*}
\xi^* =\left( \xi^i  \pt_i\right)^* = \xi^i \mg_{ij}  \rd x^j\; ,
\hspace{3mm} \theta^* =\left( \theta_i  \rd x^i\right)^* = \theta_i
\mg^{ij}
\pt_j\; . 
\eeq 
 Given a splitting $\sigma$, one can
combine these maps 
 to a bundle  involution $\tau \colon E\to E$, $\theta \oplus \xi \mapsto
 \xi^* \oplus \theta^*$,  with the obvious property $\tau^2 
\equiv 1$. To simplify the notation, we will henceforth just write
$\theta + \xi$ instead of
$\theta \oplus \xi$, because the nature of $\theta$
and $\xi$ anyway indicate the position in $E \cong T^*M \oplus TM$. The bundle
 $E$ thus decomposes into $\pm 1-$eigenvalue parts, $E=E_+ \oplus
E_-$, where $E_\pm := Ker(\tau \mp 1)$. 

\begin{proposition}\label{graph}
Any Lagrangian subbundle  is a graph of an orthogonal map $E_+\to
E_-$, which can be identified with a section $\cO \in \Gamma(O(TM))$.  
\end{proposition}
{\bf Proof.} First, let us show that $\tau$ is symmetric with respect to 
 $\langle ,\rangle$. In fact, by the definition of the
scalar product (\ref{prod1}) and $\tau$ we have
\beqn
\langle \tau (\xi_1 +\theta_1 ) , \xi_2 +\theta_2\rangle =
\langle \theta_2,  \theta_1^* \rangle + \langle\xi_1^*, \xi_2\rangle =
\mg(\xi_1 ,\xi_2 )+\mg (\theta_1 ,\theta_2 ) \, .
\eeq
Now it is easy to see that
the restriction of $\langle ,\rangle$ to $E_+$ ($E_-$) is a
 positive (negative) metric, respectively, and $\langle E_+ , E_-\rangle
\equiv 0$. Therefore we conclude that  
any Lagrangian subbundle does not intersect $E_\pm$. Hence the
projection of $D$ to $E_+$ is bijective which implies that $D$ 
 is a graph of some map $E_+\to E_-$. Let us identify $E_\pm$ with $TM$
 by means of $\pm \rho$, then the map  
uniquely corresponds to an orthogonal transformation $\cO$ of $TM$.
More precisely, any section $u_{\pm}$ 
of $E_\pm$ can be 
uniquely represented as $\xi \pm \xi^*$ for some vector field $\xi$. 
Now the definition of $\cO$ yields that any section of $D$ has the form
$(1-\cO)\xi + ((1+\cO)\xi)^*$ for a certain vector field $\xi$.  
 Taking into account that 
$\langle ,\rangle$ vanishes on $D$, we show that $\cO$
is an orthogonal map:
\beqn &&
\! \! \!\! \! \!\! \! \!\! 
{\langle (1-\cO)\xi +((1+\cO)\xi)^* ,(1-\cO)\xi' +((1+\cO)\xi')^* \rangle} =
 \\ \nonumber &&\quad = \mg((1\!-\!\cO)\xi , (1\!+\!\cO)\xi') +
 \mg((1\!+\!\cO)\xi ,(1\!-\!\cO)\xi' ) =
2(\xi ,\xi' )-2\mg(\cO\xi ,\cO\xi' )=0  \, .
\eeq  $\blacksquare$

For the argumentation above, in particular for the fact that $D$ as a
trivial intersection with $E_\pm$, it was important that $\tg$ is a metric
of definite signature. Note, however, that even when $\tg$ is an
arbitrary pseudo-Riemannian metric, we obtain a maximally isotropic
subbundle $D$ from a graph in $E_+$, $E_-$
of an pseudo-orthogonal operator $\cO$; just not all such subbundles $D$
can be characterized in this way. This is an important fact when we
want to cover, e.g., the $\, \GG$-model for non-compact semi-simple Lie
group. 
 
Locally not 
any Dirac structure $D$ admits a splitting $\sigma$  such
that $D$ corresponds to  either a Poisson or a presymplectic
structure. But even if it does so locally, there may be global
obstructions for it to be a WZ-Poisson or WZ-presymplectic.
This can be shown by constructing characteristic classes associated to
a given Dirac structure $D \subset E$ \cite{AKS}. An example for such
a Dirac structure with
``non-trivial  winding'' is the following one:\footnote{This example
can be extracted directly from the previous section, cf.~the text
after formula (\ref{GGHPoisson})---we only changed from a
right-invariant basis to a left-invariant one.}

\begin{example} \label{ex:Cartan}
Take $M=\Gr$ a Lie group whose Lie algebra $\g =Lie \, \Gr$ is quadratic,
with the non-degenerate ad-invariant scalar product denoted by
$\langle \cdot , \cdot \rangle$. 
Then the respective exact Courant algebroid $E=
T^*\Gr \oplus T\Gr$ can be cast into the following form:
\beq
\be{ccc}
E &=& \Gr \times (\g\oplus \g) \\
\rho (x,y) &=& x^R \equiv  x\gr  \\
\langle (x ,y), (x', y')\rangle &=& \langle x,y'\rangle + \langle x',y
\rangle \nonumber \\
(x,y)\circ (x',y') &=& (-[x,x'], [x,x']-[x,y'] +[x', y]), \;
\forall \mathrm{const.~sections\;} x,x',y,y' 
\ee
\eeq
The cotangent bundle $T^*\Gr$ is embedded as follows:
$\theta \mto (0, \theta^*\gr^{-1})$, where $\theta^*$ is the vector
field dual to the 1-form $\theta$ via the Killing metric on $\Gr$
which is left- and right-invariant. Note that
for any left or right invariant vector field
$\xi$ one has $L_\xi (\theta)^* = (L_\xi \theta )^*$.

It is easy to see that the curvature $H$ of the splitting
(connection)
$\sigma\colon \xi \to (\xi g^{-1}, 0)$ 
equals the Cartan 3-form
\beqn
H(\xi_1, \xi_2, \xi_3)=
\langle \xi_1\gr^{-1}, [\xi_2\gr^{-1}, \xi_3\gr^{-1}]\rangle \; ,
\eeq
for $\xi_i\in \Gamma (TG)$.
The natural Dirac structure, considered in section 2, is determined by
$\cO=Ad_\gr$. One can calculate the product of two section of this Dirac 
structure in the representation defined above (here $x,y$ are constant 
sections of $\Gr\times \g$):
\beq
\left( (1-\cO)x,(1+\cO)x\right) \circ 
\left( (1-\cO)y ,(1+\cO)y\right) =
\left(- (1-\cO)[x, y], - (1+\cO)[x, y]\right) \; .
\eeq
\end{example}

Certainly, closure on $\Gamma(D)$ of the induced product or bracket
requires some additional
property of the operator $\cO$, generalizing e.g.~the Jacobi identity
of the Poisson bivector in Example 1.
\begin{proposition} \label{prop:O}
A Lagrangian subbundle, represented by
an orthogonal operator $\cO$ as the set
$D=\{(1-\cO)\xi + ((1+\cO)\xi)^* \}$,
is a Dirac structure, iff
the following property holds,
where $\nabla$ denotes the Levi-Civita connection on $M$ and $\xi_i\in
\Gamma (M,TM)$:
\beq\label{3property}
\sum\limits_{\sigma\in \mathbb{Z}_3} \; 
\mg \left( \cO^{-1}\nabla_{(1-\cO)\xi_{\sigma(1)}}(\cO)\; \xi_{\sigma (2)} 
\;, \;
\xi_{\sigma (3)}\right)= \frac{1}{2}H\left( (1\!-\!\cO)\xi_1,
 (1\!-\!\cO)\xi_2 ,(1\!-\!\cO)\xi_3\right)
\; .
\eeq

\end{proposition}
\proof
First, we rewrite the multiplication law in terms of
 the Levi-Civita connection:
\beq\label{covmult1}
x_1\circ x_2= \na_{\rho (x_1)} x_2-\na_{\rho (x_2)}x_1 + \langle \na x_1, x_2
\rangle + H(\rho(x_1), \rho (x_2), \cdot )\; ,
\eeq
where $x, y\in E$,
$\na x$ is thought of as a 1-form taking values in $E$, and hence
$\langle \na x_1, x_2\rangle$ is in $\Omega^1 (M)\subset E$.


Let us take $x_i\in \Gamma (M, D)$, $i=1,2,3$, written in the form
\beq
x_i= (1-\cO)\xi_i + ((1+\cO)\xi_i)^* \, . \label{parametrization}
\eeq
Note  that $\rho (x_i)=
(1-\cO)\xi_i$.
 Using (\ref{covmult1}),
we derive the product $x_1\circ x_2$ and the 3-product
 $\langle x_1\circ x_2, x_3\rangle$, which is a $C^\infty
(M)-$linear form vanishing if and only if $D$ is closed with respect to the
Courant multiplication
\beqn
x_1\circ x_2 &=&
(1\!-\!\cO)\left(\na_{\rho (x_1)}\xi_2 - \na_{\rho (x_2)}\xi_1\right)
 +  \left((1\!+\!\cO)\left(\na_{\rho (x_1)}\xi_2 - \na_{\rho
 (x_2)}\xi_1\right) \right)^*   \\ \label{prod}
 &&+ \, H(\rho(x_1), \rho (x_2), \cdot )
  - \na_{\rho (x_1)} (\cO) \xi_2 + \na_{\rho
 (x_2)}(\cO)\xi_1  \\ \nonumber && + \,
 \left(   \na_{\rho (x_1)} (\cO) \xi_2 - \na_{\rho
 (x_2)}(\cO)\xi_1 \right)^* -2 \mg\left( \cO^{-1}\na (\cO) \xi_1 ,\xi_2 \right)
\; .
\eeq
In the above we used 
 that the Levi-Civita connection commutes with $\tau$, i.e.
$\na_\xi (\eta^*)=\left(\na_\xi \eta\right)^*$.
 Apparently, the sum of the first and second terms
 in (\ref{prod}) belongs to the same maximally
 isotropic subbundle, therefore its product with $x_3$ vanishes, and
 $\;\langle x_1\circ x_2, x_3\rangle = (I)+(II)+(III)$, where
 \beqn
 (I) &=& 
\langle  - \na_{\rho (x_1)} (\cO) \xi_2 + 
  \left(   \na_{\rho (x_1)} (\cO) \xi_2  \right)^*  ,  
 (1\!-\!\cO)\xi_3 +  \left( (1\!+\!\cO)\xi_3 \right)^* 
  \rangle - (1\!\leftrightarrow\! 2)=\\ \nonumber
&=&  \mg\left(   - \na_{\rho (x_1)} (\cO) \xi_2  , (1\!+\!\cO)\xi_3 \right)+ 
\mg\left( \na_{\rho (x_1)} (\cO) \xi_2
  ,(1\!-\!\cO)\xi_3 \right)- (1\!\leftrightarrow \! 2)
 = \\ \nonumber &=&
 - 2 \mg\left( \cO^{-1}\na_{\rho (x_1)} (\cO) \xi_2 + \cO^{-1}\na_{\rho
 (x_2)}(\cO)\xi_3 , \xi_1 \right)\; .
  \eeq
and
\beqn
(II) &=& \langle
- 2 \mg\left( \cO^{-1}\na (\cO) \xi_1 ,\xi_2 \right), x_3\rangle
=  - 2 \mg\left( \cO^{-1}\na_{\rho (x_3)} (\cO) \xi_1 ,\xi_2 \right)\: , \\
\nonumber
(III)&=& H\left( (1-\cO)\xi_{\sigma(1)},
 (1-\cO)\xi_{\sigma(2)} ,(1-\cO)\xi_{\sigma(3)}\right)\: . \eeq
 In the formulas above
 $ (1\!\leftrightarrow \!2)$ denotes the permutation of the first two 
indices and $ \mathbb{Z}_3$ is the group of cyclic permutations of
 order 3. We also used that the orthogonality of $\cO$ implies that
$\cO^{-1}\nabla (\cO) $ is a skew-symmetric operator with
respect to the metric $\mg$, i.e.~$\mg (\cO^{-1}\nabla (\cO)\eta_1
, \eta_2)
=- \mg (\cO^{-1}\nabla(\cO)\eta_2 , \eta_1)$ holds 
for any couple of vector fields
$\eta_1, \eta_2$. All in all we then obtain
\beq\label{3formWZW}
\langle x_1\circ x_2, x_3\rangle &=&
-2\!\sum\limits_{\sigma\in \mathbb{Z}_3} 
\mg \left( \cO^{-1}\nabla_{(1-\cO) \xi_{\sigma(1)}}(\cO),\xi_{\sigma (2)} , 
\xi_{\sigma (3)}\right) + \\ \nonumber &+&
 H \left((1-\cO) \xi_1, (1-\cO) \xi_2,
(1-\cO) \xi_3\right) \:,
\eeq 
which implies (\ref{3property}).
$\blacksquare$

{}From the above proof we extract the following useful
\begin{cor}
Assume that the integrability condition (\ref{3property}) holds and
that $x_i \in \Gamma(D)$, parameterized as in (\ref{parametrization}). 
Then
their Courant product (\ref{prod}) can be written as
\beq\label{covprod}
x_1\circ x_2 =
(1-\cO)Q(\xi_1, \xi_2) + \left((1+\cO)Q(\xi_1, \xi_2) \right)^* \; , 
\eeq
where 
\beq
Q(\xi_1, \xi_2)\!=\!\na_{\rho (x_1)}\xi_2  \! - \! \na_{\rho
(x_2)}\xi_1 \! +
 \!\left( \mg\left( \xi_1,  \cO^{-1}\na (\cO) \xi_2\right)\right)^*  \!
+ \!\frac{1}{2}  H \left(\rho(x_1),  
\rho(x_2), \cdot \right)^* , \label{Q}
\eeq
and $\rho(x_i) \equiv (1 \!- \!\cO) \xi_i$. 
\end{cor}
At the expense of introducing an auxiliary metric $\tg$ on
$M$, a Dirac structure can be described globally by  $\cO \in \Gamma(O(TM))$. 
The introduction of $\CO$ permits also to identify $D$ with
$TM$ (via Eq.~(\ref{parametrization}). The Courant bracket thus induces a
Lie algebroid bracket on $D$. This in turn induces an unorthodox
Lie algebroid
structure on $TM$, where the bracket between two vector fields $\xi$,
$\xi'$ is given by
$[\xi , \xi' ]: = Q(\xi, \xi')$, which defines a Lie algebroid
structure on $F := TM$ with anchor $\rho \colon F \to TM$,
$\xi \mapsto (1 - \CO) \xi$. In
a holonomic frame,  the corresponding structure functions, $[\partial_i
, \partial_j]_F = C^k_{ij} \partial_k$, are
easily computed as 
\beq\label{structurefunctions}
  C^k_{ij} = {(1\!-\!\cO)^m}_i
\Gamma_{mj}^k - (i \! \leftrightarrow \! j) 
+ {\cO^{m}}_{j}{}^{ ; k}\cO_{m i}
 +\frac{1}{2} H_{mn}{}^k (1\!-\!\cO)^m_i (1\!-\!\cO)^n_j  \; .
\eeq

For practical purposes it may be useful to know how the orthogonal 
operator $\cO$
transforms when changing $\tg$:
\begin{proposition} \label{prop:changem}
 Given a fixed splitting so that $E =T^*M \oplus TM$,
the couples $(\tg,\cO)$ and $(\tg,\widetilde \cO)$ describe the same 
 Dirac structure $D$ on $M$ iff
\beq \widetilde \cO = \left[\cO - 1 + \widetilde \tg^{-1} \tg (1 +
\cO) \right] \, 
\left[1 - \cO + \widetilde \tg^{-1} \tg (1 + \cO)\right]^{-1}\, .
\eeq
\end{proposition}
{\bf Proof.} 
An arbitrary element $\o \oplus v$ in $D$ can be parameterized as 
$\tg(1+\cO)\xi \oplus (1-\cO)\xi$ for some $\xi \in TM$. Equating this 
to $\widetilde \tg(1+\widetilde \cO)\widetilde \xi \oplus 
(1-\widetilde \cO)\widetilde \xi$, it is elementary to derive
$\widetilde \xi = \2 \left[1 - \cO + \widetilde \tg^{-1} 
\tg (1 + \cO)\right] \xi$. Since both of two parameterizations 
are one-to-one (see Proposition
 \ref{graph}), the
dependence above is invertible. 
 Using this relation in equating $(1-\cO)\xi$
to $(1-\widetilde \cO)\widetilde \xi$ for all $\xi$, we prove 
the statement of the proposition.
$\blacksquare$ 

As a simple corollary one obtains the following 
\begin{lemma}\label{invertibility} 
For any orthogonal operator $\cO$ and positive or
negative symmetric operator 
$b$, the  operator $1 - \cO + b (1 + \cO)$ is invertible. 
\label{lemma1}
\end{lemma}
Both assumptions in the Lemma refer to a definite metric (as this was
assumed and necessary for an exhaustive description of Dirac
structures---cf.~the discussion following Proposition \ref{graph}). 
For later use we conclude from this
\begin{cor} \label{corop} 
The operator
\beq T_{\!\a} := 1 + \cO + \alpha (1 - \cO) * \label{eq:op}\eeq
on $T^*\S \otimes X^*TM$ is invertible. Here $*$ is the Hodge
operator on $T^*\S$; for Lorentzian signatures of $h$, $*^2 = \mathrm{id}$
and, by assumption, $\alpha \in \R\! \setminus\! 0$,
for Euclidean signatures, $*^2 = -\mathrm{id}$ and $i\alpha \in
\R\!\setminus \!0$. 
\end{cor}
The statement above follows in an obvious way from Lemma \ref{lemma1},
i.e.~for definite metrics $\tg$. For pseudo-Riemannian
metrics $\tg$, however, it in general becomes necessary
to restrict $\alpha$ to a neighborhood of $\alpha = 1$ and 
$\alpha = i$  for
Lorentzian and Euclidean signature of $h$, respectively.





\section{Change of splitting}\label{sec:split}
%
The action $\SDSM$ of the Dirac Sigma Model consists of two parts, the 
topological term (\ref{Stop}) and the kinetic one (\ref{Skin}). It was mentioned 
repeatedly that only the 
second contribution depends on the auxiliary metrics $\mg$ and $h$. However, 
also the first part $\Stop$ (and in fact now only this part) 
depends on another auxiliary structure, namely the 
choice of the splitting. We will show in the present section that this dependence 
is rather mild: It can be compensated by a  
coordinate transformation 
on the field space, which is trivial on the classical solutions 
(cf.~Proposition \ref{prop:fieldredefinition} below; the
transformation is $\a$-dependent, so it 
changes if also the kinetic term is taken into account).

There is also an interesting 
alternative: Recall that $h^{-1} \otimes X^*\mg$ was used as a
symmetric pairing in $\Gamma(T^*\S 
\otimes X^* TM)$ to define $\Skin$. If in addition we are given a  2-form 
$C$ on $M$, we can also use $h^{-1}(\mathrm{vol}_\S) \otimes  X^* C$
for a symmetric pairing, 
where $\mathrm{vol}_\S$ is the volume 2-form on $\S$ induced by $h$,
and
$h^{-1}(\mathrm{vol}_\S)$ denotes 
the corresponding bivector resulting from raising indices by means of
$h$. Using the sum 
of both (or, more precisely, an $\alpha$-dependent linear combination of them) to 
define $\Skin$, cf.~Eq.~(\ref{eq:newkin}) below, 
a change of splitting can be compensated by a simple change of the new
background field $C$.

{}From Section \ref{Background} one 
knows that a splitting in an exact Courant 
algebroid is governed by 2-forms $B$. Namely, assume that $\sigma :TM\to E$
is a splitting, then any other one sends a vector field $\xi$ 
to $\sigma_{\! B} (\xi):=\sigma (\xi ) + B (\xi ,\cdot )$.

\begin{proposition} \label{prop:change}
The DSM action transforms under a change of splitting $\s \to \s_{\!
B}$ according to:
\beq\label{B-dependence}
\SDSM \; \mto \; \widetilde{S}_{\DSM} : =
\SDSM+ \frac{1}{2}\int\limits_{\Sigma} B_{ij}\, f^i \wedge f^j  \;, 
\eeq
where $f^i \equiv \md X^i - V^i$.
\end{proposition}
\proof
In fact, the decomposition $\cA = A+V$ is not unique and depends on
the 
splitting. Changing the splitting by a 2-form $B$, we get a different
decomposition:
$\cA = \widetilde{A} +V$ where $\widetilde{A} = A - B(V, \cdot )$.
To argue for this we note that $V=\sigma\left(\rho (\cA )\right)$ 
is indeed invariant (in particular, $\Skin$ is invariant), only $A$
varies, hence after the change of splitting we obtain 
$\widetilde{A}=\cA -\sigma_{\! B}\left(\rho (\cA )\right)
= A - B(V, \cdot )$. Taking into account that the $B$-field influences
on $H$, $H\mto \widetilde{H}=H+dB$, we calculate:
\beqn
\widetilde{S}_{top}&=&
\int\limits_\Sigma \widetilde{A}_i \wedge \md X^i -
\frac{1}{2} \widetilde{A}_i \wedge V^i +\int\limits_{N_3}  \widetilde{H}=
\\ \nonumber &=&
S_{top} +\frac{1}{2}\int\limits_\Sigma B(\md X \w \md X)- 2 
B(V \w \md X) + B(V \w V) \; ,
\eeq
which finally gives the required derivation (\ref{B-dependence}).
$\blacksquare$

As mentioned above, if 
the kinetic term $\Skin$ (\ref{GGSkin}) is replaced by
\beq \Skin^{\mathrm{new}} := \frac{1}{2}
\int_\Sigma 
\alpha \, \mg ( f\!  \w \!*f) +  C ( f \! \w \!
f)
\label{eq:newkin} \eeq
for some auxiliary 
$C \! \in \!\Omega^2(M)$, then a change of splitting governed by the  $B$-field merely
leads to $C\mto C-B$ for this new background field. 

Note that despite the fact that $H$ and $C$ change in a similar way w.r.t.~a 
change of splitting,  $H \mapsto H +  \md B$, $C \mapsto C - B$, they enter 
the sigma model qualitatively
in quite a different way: $H$ is already \emph{uniquely} 
given by the Courant algebroid and a chosen splitting,
while $C$ is on the same footing 
as $\mg$ or $h$, which have to be chosen in addition.
In what follows we will show that 
a change of splitting does not necessarily lead to a change in the
background fields, 
but instead can also be compensated by a transformation of the field
variable---at least infinitesimally. 

\begin{proposition} \label{prop:fieldredefinition}
Let $\a \ne 0$ and $B$ be a ``sufficiently small'' 2-form. Then there exists
a change of variables $\bar\ca :=\ca +\de\ca$ such that
$\SDSM[X,\bar\ca ]\equiv \SDSM[X,\ca ] +\frac{1}{2} \int_\S
B(f\w f)$.
\end{proposition}
\vskip-0.5em
Clearly, $\de \ca$ needs to vanish for $f=0$. We remark that this
equation is one of the field equations, cf.~Theorem \ref{theo1}
below, so that  $\de \ca$ corresponds
to an on-shell-trivial coordinate transformation (on field space).  
\proof
We find that after the change of variables
$\bar\ca :=\ca +\de\ca$ one has
\beq \label{DSMtrafo}
\SDSM[X,\bar\ca ]- \SDSM[X,\ca ] =
\int\limits_\S \tg (T_{\!\a} \,
\de\ca \w f) + \frac{1}{2}\tg (\de\ca \w R_\a \, \de \ca)
\; ,
\eeq
where $T_{\!\a}$ is given by eq.~(\ref{eq:op}) and  
 \beq
R_\a = \cO -\cO^{-1} +\a  (2-\cO -\cO^{-1})\! *  .
\eeq
Solving the equation $\SDSM[X,\bar\ca ]- \SDSM[X,\ca ] =
\frac{1}{2} \int_\S B(f\!\w\! f) $, we use the ansatz 
$\de\ca =T_{\!\a}^{-1}Lf$ for some yet undetermined $L \in
\Gamma(\S,\End (T^*\S \otimes
X^*TM))$ (invertibility of $T_{\!\a}$ follows from Corollary
\ref{corop}). This yields the following equation for $L$:
\beq L^* A L + L^* - L + B = 0 \, , \label{eq:L*}\eeq
where $A = T_\a^{{-1}*} R_\a^* T_\a^{-1}$ and $B$ denotes
the operator via the identification $B(a\w b) = \tg ( Ba \w b)$,
i.e.~the operator is obtained from the bilinear form by raising the
second index. In the above the adjoint $L^*$ of an operator $L$
in $T^*\S \otimes
X^*TM$ is defined by means of the canonical pairing induced by $\tg$:
$\tg (a \! \w \! L b) = \tg (L^* a \! \w \! b)$.\footnote{For
operators commuting with the \label{footnote*}
Hodge dual operation $*$ (which applies to all opartors appearing
here), this coincides with the adjoint defined by the symmetric
pairing induced by $\tg$ and $h$: 
$\tg (a \! \w \!* L b) = \tg (L^*a \! \w\! *b)$. One may then 
verify e.g.~$R_\a^*=-R_\a$, $*^*=-*$, $A^*=-A$, and $T_\a^*=\cO^{-1}T_\a$.
This notation is not to be confused with the isomorphism between $TM$
and $T^*M$, extended to an involution $\tau \colon
TM\oplus T^*M$, denoted by the same symbol, cf.~Eq.~(\ref{eq:*}).}
For sufficiently small $B$, the above
operator (or matrix) equation (\ref{eq:L*})
has the following solution:
\beq L = \left( \sum_{n=0}^\infty { \2 \choose n+1} \, (BA)^n  \right) \, B
\, , \label{eq:sum}
\eeq
where ${ \l \choose n} \equiv \frac{\l (\l-1) \ldots (\l - n +
1)}{n!}$. Note that the $L$ above is anti-selfadjoint (antisymmetric),
$L^* = -L$, so it solves the simplified equation $LAL + 2L -B =0$. For
the case that $A$ has an inverse, 
the
above solution can also be rewritten in the more transparent form
$L=A^{-1} \left(\sqrt{1+AB} - 1\right)$. Since $A$ can be seen to be
bounded, the sum (\ref{eq:sum})
converges for small enough $B$.
$\blacksquare$

For completeness we also display how $\cO$ transforms under a
change of splitting:
\begin{proposition} Given two splittings $\sigma ,\widetilde\sigma$ of
the exact
Courant algebroid $E$, the couples $(\s , \cO)$ and $(\widetilde\s , 
\widetilde\cO)$ describe the same Dirac structure $D$ on $M$, iff
\beq
\widetilde\cO = \left( 2\cO -B (1-\cO)\right)\left( 2-B (1-\cO)
\right)^{-1} \; ,
\eeq
where $B$ is defined as follows: $\widetilde\s (\xi )=\s (\xi)
+ (B\xi )^*$ for any vector field $\xi$.
\end{proposition}
\proof Straightforward calculations similar to Proposition \ref{prop:changem}.
$\blacksquare$

\section{Field equations and gauge symmetries}
\label{sec:eom}
In this section we compute 
the equations of motion of the Dirac sigma model (DSM) introduced in section 
\ref{sec:GG} and reinterpret them mathematically.  In particular we will show 
that the collection $(X,\CA)$ of the fields of the DSM are solutions to the 
field equations if and only if they correspond to a morphism from $T\S$ to 
the Dirac structure $D$, viewed as Lie algebroids. As a consequence they are 
also independent of the choice of metrics used to define the kinetic term 
$\Skin$ of the model as well as of the splitting used to define
$\Stop$. 

\begin{deff}[\cite{Mc,BKS}]\label{strong}
A vector bundle morphism $\phi\colon E_1 \to E_2$ 
between two Lie algebroids with the anchor maps
$\rho_i : E_i \to TM_i $ is a \emph{morphism of Lie algebroids}, iff
the induced map $\Phi \colon \Gamma (\Lambda E_2^*) \to
\Gamma (\Lambda E_1^*)$
is a chain map with respect to the canonical 
differentials $\md_i$:
\beq \label{LA_morph} \md_1 \Phi -\Phi\md_2 =0    \,. 
\eeq
\end{deff}
First, notice that, fixing a base map $X \colon M_1\to M_2$, any vector
 bundle morphism is uniquely determined by a section $a\in \Gamma
 (M_1, E_1^* \otimes X^* E_2)$. Hence the morphism property (\ref{LA_morph})
should have
 a reformulation in terms of the couple $(X,a)$. Second, it is easy to 
see that the property (\ref{LA_morph}) is purely local, therefore  it admits
a description for any local frame.

Indeed, let $\{e_i\}$ be a local frame of the vector bundle $E_2$,  
$\{e^i\}$ be its dual, and $a^i : =\Phi (e^i)$, then (\ref{LA_morph}) is 
equivalent to the following system of equations:
\beq\label{first} \md_1 X-\rho_2 (a)=0 \\ \label{second}
\md_1 a^k +\frac{1}{2} C^k_{ij}\; a^i\wedge a^j =0\eeq 
The first equation  is covariant; it implies that the 
following commutative diagram holds
true:
\beq\nonumber
E_1 & \stackrel{\phi }{\longrightarrow} & E_2 \\ \nonumber
\rho_1 \downarrow && \rho_2 \downarrow \\  \nonumber
TM_1 & \stackrel{X_* }{\longrightarrow} & TM_2
\eeq
The second equation depends on the choice of frame. However, the additional
contribution, which arises in (\ref{second}) under a change of frame, is 
proportional to 
the $\md_1 X-\rho_2 (a)$, hence it is of no effect, if the equation
(\ref{first}) holds.
For further details about this definition we refer to \cite{BKS}. 

In our context $E_1 = T\S$ and thus $\rd_1$ becomes the ordinary de
Rham differential. The $a$ above becomes $\cA \in \O^1(\S,X^*D)$, but
can be identified with $\ca$ due to 
\beq \label{CApara}
\CA &\equiv& A + V = \left((1+\CO) \ca\right)^* + (1-\CO) \ca \,
, \eeq
where $\ca \in \O^1(\S,X^*TM)$ is an unrestricted field. 
In these variables $\SDSM$ has the form (\ref{eq:DSMca}). 
The first morphism property, equation (\ref{first}), then takes the
form
$f\equiv \rd X - V =0$.

\begin{theorem}
\label{theo1}
Let $\a \ne 0$ (depending on the signatures of $h$ and $\tg$,
possibly further restricted as specified after eq.~(\ref{practDSM}) or
at the end of section \ref{Background}). Then the field equations of
$\SDSM$ have
the form
\beq
\label{1st}f \equiv \md X- (1-\cO)\ca = 0 \; , \\ \label{2d} 
 \bt (\ca)  + \frac{1}{2} \mg \left( \ca  \w \cO^{-1}\na
(\cO) \ca\right)^* +\frac{1}{4} H((1-\cO)\ca \w (1-\cO)\ca,
\cdot )^* = 0\;  ,
\eeq
or, in the dependent $(A,V)$ variables,  $\md X = V$ and 
\beq \label{2nd}
 \bt A + 
\frac{1}{4}  \mg \left(V \w\cO^{-1}\na
(\cO) V \right) + 
\frac{1}{4} \langle A \w 
\cO^{-1}\na (\cO) A^* \rangle
+ \frac{1}{2} H(V\w V\w \cdot) = 0 \, .
\eeq
The fields $(X,\cA)$ are a solution of the
equations of motion, if and only if they induce 
a Lie algebroid morphism from  $T\Sigma$ to $D$.
\end{theorem}
{\bf Corollary.} The classical solutions of the DSM do not depend on the
choice of the coupling constant $\a \ne0$ (in the permitted domain),
or, more generally, on
the choice of metrics  $\mg$ and $h$. \\

\noindent \proof
Using that $\cO$ is an orthogonal operator w.r.t.~$\tg$, one computes
in a straightforward generalization of (\ref{eom1a}) and
(\ref{eom1b})\footnote{Recall that $a = -\ca$ and $\cO = \Adg$ in the
$\,\GG$ model. The variational derivative is defined according to
$\delta_\ca \SDSM = \int_\S \langle\delta \ca \w  \frac{\delta \SDSM}{\delta \ca}
 \rangle
\equiv \int_\S \tg \left(\delta \ca
\w \left( \frac{\delta \SDSM}{\delta \ca}\right)^*
 \right)$.  
Alternatively, one may infer this relation also from
Eq.~(\ref{DSMtrafo}), keeping only the terms quadratic in $\de \ca$:
since $\tg (T_{\!\a} \,
\de\ca \w f) = \tg ( 
\de\ca \w \CO^{-1} T_{\!\a}f)$, cf.~footnote \ref{footnote*}.}
\beq \CO \, \left(\frac{\delta \SDSM}{\delta \ca} \right)^* =
\left( 1 + \cO + \a (1 - \CO) * \right) f
\, . \label{eq:avar}
\eeq 
The term in the brackets is the operator (\ref{eq:op}), which,
according to our assumption on $\a$, is invertible;
so, indeed  $f = 0$. Note, however, that in general $1 + \CO$ is
invertible only if the Dirac structure corresponds to a graph of a
bivector. Thus, \emph{only} in the WZ-Poisson case one may drop the
kinetic term altogether \emph{if} one wants to keep the morphism
property of the field equations.

We now turn to the $X$-variation of $\SDSM$. This is conceptually more
subtle since $\ca \in \Omega^1(\Sigma, X^*TM)$ depends implicitly on
$X$ as well. Thus to determine
$\delta_X a$ we need a connection, since, heuristically,
we are comparing sections in two different, but nearby bundles $X_0^*
TM$ and $(X_0 + \delta X)^*TM$. (If we required e.g.~$\delta_X \ca^i=0$,
then this would single out a particular holonomic frame, since a change of
coordinates on $M$ yields $\widetilde \ca^i = M^i_j(X) \ca^j$. In the
following we develop an inherently covariant formalism that also produces
covariant field equations.)

Let us use denote the local basis in $X^* TM$ dual to $\bard
X^i$ in $X^*T^*M$ by $\barp_i$. The notation $\bard
X^i$ is used so as to distinguish it from
$\rd$ acting on the pull-back function
$X^*X^i$, which we denote as usual by $\rd X^i$. Then
\beq \label{delta}
\delta_X \barp_i = \Gamma^j_{ki}\delta X^k \barp_j \, , \eeq
where $\Gamma^i_{jk}$ are coefficients of the Levi Civita
connection $\nabla$ of $\tg$. Also, we think of
$\ca^i$ to depend on both $X(x)$ and $x$ (cf.~also \cite{BKS} for
further details); correspondingly, $\delta_X \ca = \left(\delta_X (\ca^i) \,
+ \Gamma^i_{kj}\, \ca^j\delta X^k \right) \barp_i$.
 Note that
certainly $\delta_X \rd = \rd \delta_X$, where $\rd$ denotes the de
Rahm operator. However, this does not apply for  $\rd X$
used above, which 
  is the section in $T^*\S \otimes X^*TM$ corresponding to
the bundle map $X_* \colon T\S \to TM$, $\rd X = \rd X^i \otimes
\partial_i$; so in $\rd X$, $\rd$ does not denote an operator. Here
one finds in analogy to $\delta_X a$
\beq
\delta_{X} (\md X) =\left( \md (\de X^i) + \Gamma^i_{kj} \md X^j \de X^k
\right) \; \barp_i =\bt (\de X) \; , \label{eq:deltadX}
\eeq
where $\bt$ is the pull-back of the Levi-Civita connection $\na$ to $X^*
(TM)$ and in the last step torsion freeness  of $\na$,
$\Gamma^i_{kj}= \Gamma^i_{jk}$, was used. 
The above covariant form of the variation (\ref{delta}) 
 implies in
particular that $\delta_X \tg = 0$, when $\tg$ is viewed as an element
in $\Gamma(\S,X^* T^*M^{\otimes 2})$---as it appears in the action
functional (\ref{practDSM}), so we will be permitted to use
$\de_X \mg (a\! \w \!b) = \mg (\de_X a\! \w \! b) +
\mg(a \!\w \! \de_X b)$ below.

With the above machinery at hand, the variation w.r.t.~$X$ is rather
straightforward again. By construction it will produce a covariant
form of the respective field equations. Since we already know that
$f=0$ holds true, moreover, we will be permitted to drop all terms
below which are proportional to $f$. Correspondingly, 
$\Skin$ can be dropped for the
calculation of $\de_X \SDSM=0$, since $\Skin$ is quadratic in
$f$. By convention, we will denote
equalities up to $f=0$ in what follows by $\approx$; so, in particular
\beq \frac{\delta \SDSM}{\delta X} \approx 
 \frac{\delta \Stop}{\delta X} \, . \eeq
 Also, we may drop all terms containing $\delta_X \ca$, since on
behalf of (\ref{eq:avar}), they
will be proportional to $f$ again, with  or without
the kinetic term included (corresponding only to
$\alpha \neq 0$ and $\alpha =0$, respectively, in formula  (\ref{eq:avar})). 
Thus, 
\beq 
\delta_X \Stop &\approx & \int\limits_\Sigma \!\mg\left( (\delta_X \CO
)\ca \w \md X \right) + \mg\left((1+\cO)\ca \w \bt \delta X \right)
\nonumber \\
&& +  \mg\left(  \ca \w (\delta_X \CO) \ca\right) + 
\frac{1}{2} 
H_{ijk}\;\md X^i \wedge \md X^j 
\, \delta X^k \, , \label{eq:1}
\eeq
where we already made use of equation (\ref{eq:deltadX}). In the
second term we perform a partial integration (dropping eventual
boundary terms) and observe that
\beq \label{bta}
\bt \left((1+\cO) \ca \right) \approx 2 \! \bt \! \ca \, , 
\eeq
which follows from $2 \ca \approx (1+\cO) \ca + \md X$ and 
$\;\bt (\md X) = -\md X^i \wedge \md X^j \; \Gamma^k_{ji} \otimes
\barp_k\equiv 0 $. Replacing $\md X$ by $\ca - \CO
\ca$ in the first term, we then obtain
\beq
\delta_X \Stop \approx  \int\limits_\Sigma \!\mg\left( \ca \w 
(\cO^{-1}\delta_X \CO )\ca \right)
+ \mg\left(2 \!\bt \!\ca \w \delta X \right)
+ \frac{1}{2} 
H_{ijk}\;\md X^i \wedge \md X^j 
\, \delta X^k \, , \label{eq:2}
\eeq
where the first and the third term in (\ref{eq:1}) combined into the first
term above. This proves eq.~(\ref{2d}). Equivalence with
eq.~(\ref{2nd}) is established easily as follows: For the first term
we read eq.~(\ref{bta}) from
right to left and use $A^* = (1 + \CO) \ca$. For the second term of
(\ref{2d}) we
replace  $\ca$ by $\frac{1}{2} (A^* + V)$ and utilize the antisymmetry
of $\cO^{-1} \na  \cO$ to cancel off-diagonal terms. For the third one
we use $V = (1-\cO)\ca$. 

This leaves us with proving the equivalence of (\ref{2d}) to the
second morphism property (\ref{second}), specialized to the present
setting, where, again, $f=0$ may be used freely. So, we need to show
that (\ref{2d}) can be replaced by $\md \ca^i + \2 C^i_{jk} \ca^j
\wedge \ca^k \approx 0$, where the structure functions are given by
Eq.~(\ref{structurefunctions}). 
Since $C^i_{jk} \ca^j
\wedge \ca^k \otimes   \barp_i\equiv  Q(\barp_{\! j} , \barp_k) \, \ca^j
\wedge \ca^k $, most of the terms in (\ref{2d}) are identified easily,
and it only remains to show that
\beq
\bt \ca \equiv \left(
\md \ca^i + \Gamma^i_{jk} \md X^j \we \ca^k \right) \otimes \barp_i
\approx  \left(\md \ca^i +  \ca^j \!\wedge\! \ca^k {(1\!-\!\cO)^m}_j
\Gamma_{mk}^i \right) \otimes \barp_i \, ,\eeq
which is an obvious identity. 
$\blacksquare$

\vskip 2mm

Having established that the field equations enforce a Lie algebroid
morphism $T\S \to D$, it is natural to expect that on solutions the
gauge symmetries correspond to a homotopy of such morphisms
\cite{BKS}. This is indeed the case. For the gauge invariance of an
action functional, however, an off-shell
(and preferably global) definition of the symmetries are needed,
which is a somewhat more subtle question. In the present paper we only
provide the result of such an analysis, deferring for a derivation and
further details,  presented as an example
within a more general framework, to \cite{AT1}. 

\begin{proposition} For nonvanishing  $\a$ the infinitesimal gauge
symmetries of $\SDSM$ can be expressed in the following form
\beq\label{DSMgauge-prelim}
\de_{\e} X &=& (1-\cO)\e \; , \\ \nonumber 
\de_{\e} \ca &=& \bt (\e) -\mg \left( \cO^{-1}\na (\cO )\ca, \e \right)^* +
\frac{1}{2} H ((1-\cO)\ca, (1-\cO) \e, \cdot )^* \nonumber
\\ && +\,T_\a^{-1}\left( 
\frac{1}{2} H(f, (1-\cO)\e , \cdot )^* +(1-\a *) \na_{\!f} (\cO )\e +
M f \right) \; ,
\eeq
where $\e \in \Gamma(\S,X^*TM)$ 
and $M=M^* \in \Gamma(\End(T^*\S \otimes X^*TM))$ may be chosen freely.
\label{prop:gaugesymmetry}
\end{proposition}
The operator $M$ above parametrizes trivial gauge symmetries. In the
$G/G$ and in the Poisson case the above symmetries reproduce the known
ones for $\a =1$ and $\a = 0$, respectively. In general, however,
the inverse of $T_\a$ is defined only for nonvanishing $\a$,
cf.~Corollary \ref{corop}. 


\section{Hamiltonian formulation}\label{sec:Hamilton}
In this section we derive the constraints of the DSM. 
For simplicity 
we restrict ourselves to closed strings, $\Sigma \cong S^1 \times
\R \ni (\s,\tau)$. Here $\s \sim \s + 2 \pi$ is the ``spatial''
variable around the circle $S^1$ (along the closed string) and $\tau$
is the ``time'' variable governing the Hamiltonian evolution. 

The discussion
will be carried out for more general actions in fact: We may regard 
any action of the form of  $\SDSM$ where $D$ is required to be a
maximally isotropic (but possibly non-involutive) subbundle of $E$;
in other words for the present purpose 
we will consider any action of the form (\ref{practDSM}) for
\emph{any} orthogonal ($X$-dependent) matrix $\cO^i_j$. Generalizing
an old  fact for PSMs \cite{PSM0}, 
the corresponding constraints are ``first
class'' (define a  coisotropic submanifold in the phase space), iff 
$D$ is a Dirac structure (i.e.~iff the matrices $\cO^i_j$ satisfy the 
integrability conditions (\ref{U}) found above). 

Let
$\partial$ denote the derivative with respect to $\sigma$ (the
$\tau$-derivative will be denoted by an overdot below) and let
$\delta$ be the exterior differential on phase space. Then we have
\begin{theorem} \label{theo:Ham}
For $\a \ne 0$ and $\Sigma \cong S^1 \times \R$,
the phase space of  $S_{DSM}$, $D$ maximally isotropic in $E=T^*M
\oplus TM$, may be identified with the cotangent
bundle to the loop space in $M$ with the symplectic form twisted by
the closed 3-form $H$,
\beq \Omega = \oint_{S^1}  \delta X^i(\s) \wedge \delta
p_{i}(\s) d\s - \2  \oint_{S^1}
H_{ijk}(X(\s)) \, \partial X^i
(\s) \, \delta X^j(\s) \wedge \delta X^k(\s) \, d\s \; , \label{symp} \eeq
subject to the  constraint $J_{\omega,v}(\sigma) = v^i(\s,X(\sigma)) 
p_i(\sigma) +
\omega_i(\s,X(\s)) \partial X^i(\sigma) = 0 \,$,
for any choice of $\omega \oplus v \in C^\infty(S^1) \otimes 
\Gamma(X^*D)$,
or, in the description of $D$ by means of $\cO \in \Gamma(O(TM))$, 
\beq J \equiv (\cO + 1) \partial X + (\cO -1) p = 0 \, .\label{constraint} \eeq
The constraints are of the first class, if and only if $D$ is a Dirac structure.
\end{theorem}
For Dirac structures $D$ that 
 may be written as the graph of a bivector $\cP$ (for the 
splitting chosen---cf.~Proposition \ref{graph} and Example \ref{ex1}), 
$1 + \cO$ is invertible; then obviously 
(\ref{constraint}) can be 
rewritten as $\partial X - \cP p = 0$ or $\partial X^i + \cP^{ij} p_j = 0$ 
(cf.~Eq.~(\ref{Cayley}) and the 
text about notation following this equation!), which agrees with the 
well-known 
expression of the constraints in the WZ-Poisson sigma model \cite{Ctirad}.  

{\bf Proof.} 
To derive the Hamiltonian structure we follow the shortcut version of
Diracs procedure advocated in \cite{Faddeev-Jackiw}. For simplicity we
first drop the WZ-term,  manipulating $\int_\S \cL \md \sigma
\wedge \md \tau := \SDSM -
\int_{N_3} H$ in a first step.
With $\md x^- \wedge \md x^+ = -2 \md \sigma \wedge \md \tau$ we
obtain from $\SDSM$ by a straightforward calculation
\beq \cL[X,\CA_\pm] &=& -\frac{\a}{2} \dot X^2 + \2 (A_+ + \a V_+ - 
A_- + \a V_-)
\dot X \nonumber \\ &&+ \frac{\a}{2} \partial X^2 + \2 (-A_+ - \a V_
+ -A_- + \a V_-)
\partial X  \nonumber \\ && - \2 V_- (A_+ + \a V_+)  + \frac{1}{4}
(A_+ V_- + A_- V_+) \, , \label{cL} \eeq
where appropriate target index contraction is understood [canonically, 
 $V$, $\dot X$, and $\partial X$ carry an upper target-index, and 
$A_\pm$ 
(as well as $p$ introduced below) 
 a lower one; but 
all indices may be raised and lowered by means of the target metric $g$]. 
For what
concerns the determination of momenta, $\a \neq 0$ is
qualitatively quite different from  $\a = 0$. Restricting to the first 
case for the proof of the present theorem, we may
now employ the  following substitution to introduce a new momentum
field $p$: 
\beq \label{eq:momenta}
 -\frac{\a}{2} \dot X^2  + \beta \dot X  \quad \simto
 \quad p \dot X + \frac{1}{2\a} \left( p - \beta \right)^2 \, , 
\eeq
which results from $ -\frac{\a}{2} \dot X^2   \simto
 p \dot X + \frac{1}{2\a} p^2$ after shifting $p$ to  $p -
 \beta$. Within an action functional any such two expressions---for 
 arbitrary C-numbers $\a\neq 0$ and possibly field dependent
 functions $\b$---are
 equivalent, classically (eliminate $p$ by its field equations)
 or on the quantum level (Gaussian path integration over $p$). 
Applying this to the first line in (\ref{cL}) with $\beta = 
\frac{1}{2}(A_+ + \a V_+ - A_- + \a V_-)$ and noting that the last 
bracket in the third line vanishes due to $\cA_\pm \in D$ and the
isotropy condition posed on $D$, we obtain $\cL \simto \LH$ with  
\beq \LH[X,\CA_\pm,p] &=&  p \dot X + \frac{1}{2\a} \left( p -
\2 A_+ -\2  \a V_+ + \2 A_- -\2  \a V_- \right)^2 
 \nonumber \\ &&+ \frac{\a}{2} \partial X^2 + \2 (-A_+ - \a V_+ -A_- + \a V_-)
\partial X  \nonumber \\ && - \2 V_- (A_+ + \a V_+)   \, . \label{LH1} \eeq
It is straightforward to check that the above terms can be reassembled
such that 
\beq \LH[X,\CA_\pm,p] &=&  p \dot X - V_- p - A_- \partial X
\nonumber \\ && + \frac{1}{8\a} \left[ A_+ + \a V_+ - (A_- + \a V_-) -
  2(p + \a \partial X)\right]^2 
 \nonumber \\ && - \2 A_-V_- - p \partial X  \, . \label{LH2} \eeq
We now want to show that the last two lines may be dropped in this
expression. Here we have to be careful to take into account that $A$
and $V$ are in general not independent fields, but subject to the
restriction that their collection $\CA=A\oplus V$ lies in the
isotropic subbundle $D$. First we note that $A_-V_- \equiv \2 
\langle \CA_- ,\CA_- \rangle = 0$ due to 
$\CA_- \in D$. Next, with a shift $\cA_+ \to \widetilde \cA_+ := 
\cA_+ +  \cA_-$, the 
$\cA_-$-part drops out in the second line; this is particularly obvious  
in terms of independent fields $\ca_\pm$, where  
the term in the square brackets takes the form
$\left[(1+\cO) + \a (1-\cO) \right](\ca_+ - \ca_-)  -
  2(p + \a \partial X)$. After this shift, $\cA_-$ enters the action 
only linearly anymore, and thus plays the role of a Lagrange multiplier. 
This already shows the appearance of the constraints $J_{\omega,v}=0$.   
Parameterizing  $\omega \oplus v \in D$ as $(1+\CO) \lambda \oplus 
(1-\CO) \lambda$, one obtains $J_{\omega,v} = \mg(\CO \lambda , (\CO+1) 
\partial X +  (\CO-1) p)$, $\mg$ being the Riemann metric, which 
vanishes for any $\lambda$  
(an unconstrained Lagrange multiplier field),  
iff (\ref{constraint}) holds true. 

To show that the remaining dependence of the lower two lines in (\ref{LH2}) 
on $p \oplus \partial X \in E$ can be eliminated (by a further shift in 
the fields), is seen most easily in a path-integral type of 
argument\footnote{A purely Lagrangian argument will be provided in the 
subsequent paragraph.}: 
Integrating over $\CA_-$ (i.e.~taking the path integral over $\lambda$), one 
obtains a delta function that constrains  $p \oplus \partial X$ to lie in 
the Dirac structure $D$; correspondingly, the term $p \pX$ gives no 
contribution  since $D$ is isotropic, and $-2(p + \a \pX)$ can be absorbed 
into $\widetilde A_+ + \a \widetilde V_+$ by a further  
redefinition of $\widetilde \CA_+$ into  $\bar{\CA_+}$. After these 
manipulations the last two 
lines reduce to $\frac{1}{8\alpha} 
\left(\left[(1+\cO) + \a (1-\cO) \right] \bar \ca_+ \right)^2$. 
This is the only dependence of the resulting action on $\bar \ca_+$, which thus 
may be put to zero as well.\footnote{Note that on behalf of the
permitted values for $\a$ the the matrix 
$\a+1  + (\a -1) \cO$ is non-degenerate---due to Lemma
\ref{invertibility}
or Corollary
\ref{corop}. So the above statement follows from the field equations
of $\bar a_+$, and using 
$\bar a_+ =0$ in the action is a permitted step in the procedure of 
\cite{Faddeev-Jackiw}. However, even if $\tg$ has indefinite
signature and $\a \neq 0$ is chosen such that the above quadratic form
for $\bar \ca_+$ is degenerate, this contribution can be
dropped. Since then the action does not depend on 
directions of $\bar \ca_+$ in the kernel of the matrix, so they also give no 
contribution to the action (alternatively, $\bar \ca_+ =0$ may be 
viewed as a gauge fixation for those directions then). We remark in 
parenthesis that the rank 
of the matrix may depend on $X \in M$ in this case, but 
dropping the contribution to the action in question 
obviously is the right step.\label{footnoterank}}

There also exists an argument on the purely classical level for the 
above consideration: Denote  $p \oplus \partial X$, 
taking values in  $E$, by $\psi$, and 
$\CA_-$ and $\widetilde \CA_+ \equiv \CA_++\CA_-$ by $\lambda_D$ and 
$\mu_D$, respectively (the index $D$ so as to stress the restriction to 
the subbundle $D < E$). Then $\LH = \LH[X,p,\lambda_D ,\mu_D] = 
p \dot X - \langle \lambda_D , \psi \rangle + f_1(\mu_D - \psi) + f_2(\psi)$, 
where $f_1$ and $f_2$ are polynomial functions to be read off from 
(\ref{LH2}) and, as always, 
$\langle \cdot , \cdot \rangle$ denotes the fiber metric 
in $E$. Next we observe that $\tau(D)= \{ (v ; -\CO v) \}\cong D^*$ has 
a trivial 
intersection with $D= \{ (v ; \CO v) \}$ and $E = D \oplus \tau(D)$; 
note that $\tau(D)$ is also isotropic by construction, but in general 
will not be a Dirac structure (cf.~Eq.~(\ref{3property})); as 
indicated already by the notation, it can be identified with the dual $D^*$ 
of $D$ by means of $\langle \cdot , \cdot \rangle$. 
Thus $\psi$ can be decomposed uniquely into components 
$\psi_D \in D$ and $\psi_D^* \in D^*$, $\psi = \psi_D + \psi_D^*$. 
With $\widetilde \mu_D := 
 \mu_D - \psi_D$ the action takes the form $\LH = 
p \dot X - \langle \lambda_D , \psi_D^* 
\rangle + f_1(\widetilde \mu_D - \psi_D^*) + f_2(\psi)$. Since for vanishing 
$\psi_D^*$ the last two contributions reduce to $f_1(\widetilde \mu_D)$. 
As a consequence there exists $F(\psi, \widetilde \mu_D)$ with values in $E$ 
such that $f_1(\widetilde \mu_D - \psi_D^*) + f_2(\psi) = \langle F, 
\psi_D^* \rangle$. With $F=F_D + F_D^*$ and due to the isotropy of $D^*$ we 
then obtain  $\LH = \LH(X,p,\mu_D,\widetilde \lambda_D) = 
p \dot X - \langle \widetilde \lambda_D , \psi_D^* 
\rangle + f_1(\widetilde \mu_D)$, where 
$\widetilde \lambda_D :=  \lambda_D - F_D$ has been introduced and $f_1$ is 
a quadratic function in its argument. As before we thus may drop 
$f_1(\widetilde \mu_D)$ (cf.~also footnote \ref{footnoterank}),
obtaining 
\beq \LH = \LH[X,p,\CA_+,\CA_-] \simto 
\LH[X,p,\widetilde \lambda] = p \dot X - \mg(\widetilde \lambda, J) 
\label{LHam}
\eeq  
for some 
unconstrained Lagrange multiplier field $\widetilde \lambda \in TM$. 

Noting that the addition of the Wess-Zumino term only contributes into the 
symplectic form as specified in (\ref{symp}), we thus proved the main part
of the theorem. 

The statement about the first class property follows from specializing the 
results of 
\cite{AS}, where a Hamiltonian system with symplectic form 
(\ref{symp}) and currents $J_{\omega,v}$ for an arbitrary subset of 
elements $\omega \oplus v \in E$ was considered. 
$\blacksquare$

The constraint algebra of $J_{\omega,v}=0$ in the above theorem 
is an example of the more general current
algebra corresponding to an exact Courant algebroid $E$ found in
\cite{AS}---the first class property is tantamount to 
requiring a closed constraint or current algebra without anomalies. 
On the other hand, apparently the action 
$S_{DSM}$ 
provides a covariant action functional that produces the above
mentioned currents (as constraints or symmetry generators) 
for the case of an arbitrary Dirac structure.
As shown above this is even true if $D$ is maximally isotropic but possibly 
not a Dirac structure. It is an interesting open problem 
to consider the Hamiltonian structure of the action  (\ref{DSM}) 
for a non-isotropic 
choice of $D$ (does the more general statement hold true that the 
functional $\SDSM$ defines a topological theory iff $D \subset E$ is a Dirac 
structure?) or likewise to provide some other covariant action 
functional producing 
constraints of the form considered in  \cite{AS} for arbitrary 
$D<E$.


We already observed above that the discussion of the Hamiltonian structure 
changes (and in fact becomes somewhat more intricate) for the case of 
vanishing coupling constant $\a$. E.q.~substitutions such as in 
Eq.~(\ref{eq:momenta}) are illegitimate in this limit. Moreover, as observed 
already in previous sections of the paper,
the kinetic term $\Skin$ is even necessary in general to guarantee the 
morphism property of the field equations (it becomes superfluous only when 
$D$ is the graph of a bivector); in fact, the number of independent field 
equations may even change from $\a \neq 0$ to $\a =0$ (with the extreme case 
of $D=TM$ where for $\a=0$ one obtains no equations at all). Thus it is 
comforting to find 
\begin{theorem}
For $\a = 0$ and $\Sigma \cong S^1 \times \R$,
the Hamiltonian structure 
of  $S_{DSM}$, $D<E$ maximally isotropic, 
may be identified with the one found in Theorem \ref{theo:Ham} 
above. 
\label{theo:equiv}
\end{theorem}
{\bf Proof.} 
For a first orientation we check the statement for $D=TM$ (and $H=0$): 
Then $\SDSM \equiv 0$. The vanishing action $S$ 
(depending on whatsoever fields) 
is obviously equivalent (as a Hamiltonian system)
to an action
$\bar S[X,p,\l] = \oint ( p \dot X - \l p )$ (multiplication of the 
integrand by $\md \sigma \wedge \md \tau$ here and below is understood). 
Since this 
example of $D$ corresponds to $\CO = -1$, this latter formulation obviously 
agrees with what one finds 
in Theorem \ref{theo:Ham} for this particular case. This case already 
illustrates that a transition from $S \equiv 0$ to $\bar S$ depending on 
additional fields as written above is an important step in establishing 
the equivalence. 

We now turn to the general case, putting $H$ to zero in a first step 
as in the proof of Theorem  \ref{theo:Ham}. Thus we need to analyse 
$S_0[X,\CA] = \int_\S A_i \wedge \rd X^i - \2  A_i \wedge V^i$, with 
$\CA=A \oplus V$  taking values in $D$.
Using the unconstrained 
field $\ca \equiv \ca_0 \md \tau + \ca_1 \md \sigma$
of (\ref{CApara}), one finds 
\beq \bar S_0[X,p,\ca_0,\ca_1,\l] &=& \oint \left( p \dot X 
- \mg(\l,p - (1+\cO) \ca_1 ) \right. 
\\\nonumber && \left.
- \mg (\ca_0 , (1+\cO^{-1}) \pX  + 
( \cO -\cO^{-1}  ) \ca_1
) \right) \, .
\eeq
Here the third term just collects all terms proportional  to 
$\ca_0$ of $S_0$, as one may show by a straightforward calculation 
(using the orthogonality of $\cO$). 
The first two terms result from  
$\oint \mg ( (1+ \cO)\ca_1 , \dot X )$, the only appearance of
$\tau$-derivatives in $S_0$. The transition from $S_0$ to $\bar S_0$ is the 
obvious generalization of the analogous 
step from $S \equiv 0$ to $\bar S$ mentioned above and explains the 
appearance of the new fields $p$ and $\l$; eliminating these fields one 
obviously gets back $S_0$. 
Next we shift $\l$ 
according to $\l = \bar \l  + (1 - \cO) \ca_0$. This yields
\beq   \bar S_0[X,p,\ca_0,\ca_1,\bar \l] &=& \oint \left( p \dot X   
- \mg(\bar \l,p - (1+\cO) \ca_1 )  \right. \nonumber \\ && \left. 
- \mg (\ca_0 , (1+\cO^{-1}) \pX  + (1 - \cO^{-1}) p
) \right) \, . \label{barS}
\eeq
Note that the last term is already of the form $- \mg (\ca_0, 
\cO^{-1} J)$, with   $J$ given by Eq.~(\ref{constraint}). We now will 
argue that the second term, containing the fields 
$\ca_1$ and $\bar \l$, can be dropped. 
For the case that $1 + \cO$ is invertible, 
this is immediate since then the quadratic form for $\bar \l$ and 
$\ca_1$ in (\ref{barS}) 
is nondegenerate---$\bar \l$ and $\ca_1$ become completely determined 
by the remaining fields, without constraining them 
(cf.~also \cite{Faddeev-Jackiw}). Otherwise the variation w.r.t.~$\ca_1$ 
constrains the momentum $p$, but this constraint is fulfilled automatically 
by $J=0$, resulting from the variation w.r.t.~$\ca_0$. To turn the last 
argument into an honest off-shell argument, we perform another shift of 
variables: With $\ca_0 := \cO^{-1} \widetilde \l - \2 \l$ and $\ca_1 := 
\bar \ca_1 + \2  \cO^{-1} (p - \pX)$, Eq.~(\ref{barS}) becomes 
\beq   \bar S_0[X,p, \widetilde \l, \bar \ca_1,\bar \l] = 
\oint \left( p \dot X 
- \mg (\widetilde \l , J)
+ \mg(\bar \l,(1+\cO) \bar \ca_1 ) \right) \, . \label{barS2}
\eeq
Now it is completely obvious that the last term, the only appearance of 
$\bar \l$ and $\bar \ca_1$, can be dropped. 
This concludes our proof, since the remaining integrand  
agrees with $\LH$ in (\ref{LHam}). 
$\blacksquare$ 

As a rather immediate but important consequence the above results 
imply 
\begin{theorem} \label{independence} 
The reduced phase space of $\SDSM$ (for $\S = S^1 \times \R$ and 
$D$ any maximally isotropic $D < E$) 
does not depend on the choice of $\a \in \R$, the metrics $h$ and 
$\mg$ on $\S$ and $M$, respectively, or the splitting $\s \colon TM \to E$. 
It only depends on the subbundle $D$. 
\end{theorem}
{\bf Proof.} Independence of $\a$ follows from the above two theorems. 
Independence of $h$ is obvious and likewise the one of 
$\mg$ when the constraints are written as $J_{\o,v}(\s) =0, \; 
\forall \o \oplus v \in D$, cf.~Theorem \ref{theo:Ham} above. 
Independence on the choice of the splitting is not so obvious at first 
sight:  the symplectic form 
(\ref{symp}) depends on $H$ (and not only on its cohomology class), and 
so implicitly on the splitting, cf.~Eq.~(\ref{Hdef}); but 
likewise do the constraint functions $J_{\o,v}(\s) = \o_i \, \pX^i + v^i p_i$, 
since the presentation of an element of $D < E$ as 
$\o \oplus v \in T^*M \oplus TM$ assumes an embedding of $TM$ into $E$ (while 
$T^*M < E$ can be identified canonically as the kernel of $\rho^*$, 
cf.~Eq.~(\ref{exact}) as well as our discussion in Section
\ref{sec:split} above). The coordinate
transformation on phase space $p_i \mapsto p_i + B_{ij}\, \pX^j$
establishes the isomorphism between the two Hamiltonian structures
corresponding to different splittings (cf.~also \cite{AS}).  
Alternatively 
we may infer splitting independence also from
Propositions \ref{prop:change} and \ref{prop:fieldredefinition} above.
$\blacksquare$ 

\section*{Acknowledgement}
We are grateful to A.~Alekseev for discussions and collaboration
on related issues, from which we profited also for the purposes of the
present paper.



\end{document}